\documentclass{article}
\usepackage{amsmath}
\usepackage{graphicx}
\usepackage{color}
\usepackage{fullpage}
\newcommand{\NB}{\mathop{\mathrm{NB}}}
\newcommand{\order}{\mathcal{O}}
\newcommand{\ave}[1]{\left \langle #1 \right \rangle}

\newcommand{\diff}[2]{\frac{\mathrm{d} #1}{\mathrm{d} #2}}
\newcommand{\Ro}{\ensuremath{\mathcal{R}_0}}

\title{Edge-Based Compartmental Modeling for Infectious Disease Spread
  Part I: An Overview}
\author{Joel C. Miller\footnote{Center for Communicable Disease Dynamics, Department of Epidemiology, Harvard School of Public Health}~\footnote{Fogarty International Center, NIH}\and Anja C. Slim
\and Erik M. Volz\footnote{Department of Epidemiology, University of Michigan, Ann Arbor}}

\begin{document}

\maketitle

\begin{abstract}
The primary tool for predicting infectious disease spread and intervention effectiveness is the mass action Susceptible-Infected-Recovered model of Kermack and McKendrick~\cite{kermack}.  Its usefulness derives largely from its conceptual and mathematical simplicity; however, it incorrectly assumes all individuals have the same contact rate and contacts are fleeting.  This paper is the first of three investigating \emph{edge-based compartmental modeling}, a technique eliminating these assumptions.  In this paper, we derive simple ordinary differential equation models capturing social heterogeneity (heterogeneous contact rates) while explicitly considering the impact of contact duration.  We introduce a graphical interpretation allowing for easy derivation and communication of the model.  This paper focuses on the technique and how to apply it in different contexts.  The companion papers investigate choosing the appropriate level of complexity for a model and how to apply edge-based compartmental modeling to populations with various sub-structures.
\end{abstract}

\section{Introduction}
The conceptual and mathematical simplicity of Kermack and McKendrick's~\cite{kermack,andersonmay} Mass Action Susceptible-Infected-Recovered (SIR) model has made it the most popular quantitative tool to study infectious disease spread for over 80 years.  However, it ignores important details of the fabric of social interactions, assuming homogeneous contact rates and negligible contact duration.  Improvements are largely \emph{ad hoc}. spanning the range between mild modifications of the model and elaborate agent-based simulations~\cite{andersonmay,lloyd:computer_people,kiss:sheep,rohani:pertussis_network,eubank:episims,germann:epicast}.  Increased complexity allows us to incorporate more realistic effects, but at a price.  It becomes difficult to identify which variables drive disease spread or to address sensitivity to changing the underlying assumptions.  In this paper we show that shifting our attention to the status of an average contact rather than an average individual yields a surprisingly simple mathematical description, expanding the universe of analytically tractable models.  This allows epidemiologists to consider more realistic social interactions and test sensitivity to assumptions, improving the robustness of public health recommendations.

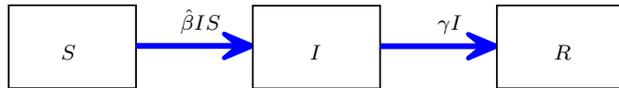
\begin{figure}
\begin{center}
\begin{picture}(0,0)%
\includegraphics{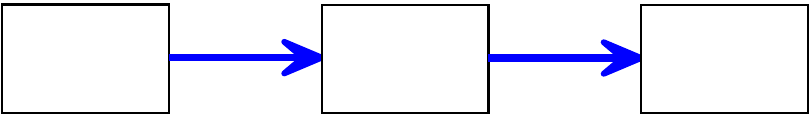}%
\end{picture}%
\setlength{\unitlength}{3947sp}%
\begingroup\makeatletter\ifx\SetFigFontNFSS\undefined%
\gdef\SetFigFontNFSS#1#2#3#4#5{%
  \reset@font\fontsize{#1}{#2pt}%
  \fontfamily{#3}\fontseries{#4}\fontshape{#5}%
  \selectfont}%
\fi\endgroup%
\begin{picture}(3883,550)(1651,-1886)
\put(2038,-1700){\makebox(0,0)[b]{\smash{{\SetFigFontNFSS{8}{9.6}{\familydefault}{\mddefault}{\updefault}{\color[rgb]{0,0,0}$S$}%
}}}}
\put(3589,-1700){\makebox(0,0)[b]{\smash{{\SetFigFontNFSS{8}{9.6}{\familydefault}{\mddefault}{\updefault}{\color[rgb]{0,0,0}$I$}%
}}}}
\put(5140,-1700){\makebox(0,0)[b]{\smash{{\SetFigFontNFSS{8}{9.6}{\familydefault}{\mddefault}{\updefault}{\color[rgb]{0,0,0}$R$}%
}}}}
\put(4432,-1509){\makebox(0,0)[b]{\smash{{\SetFigFontNFSS{8}{9.6}{\familydefault}{\mddefault}{\updefault}{\color[rgb]{0,0,0}$\gamma I$}%
}}}}
\put(2867,-1509){\makebox(0,0)[b]{\smash{{\SetFigFontNFSS{8}{9.6}{\familydefault}{\mddefault}{\updefault}{\color[rgb]{0,0,0}$\hat{\beta} I S$}%
}}}}
\end{picture}%
\end{center}
\caption{\footnotesize \textbf{Mass action flow diagram.} The flux of individuals from Susceptible to Infected to Recovered for the standard MA model.  Each compartment accumulates and loses probability at the rates given on the arrows.}
\label{fig:compartments}
\end{figure}

We motivate our approach using the standard Mass Action (MA) SIR model.  We are interested in the susceptible $S(t)$, infected $I(t)$, and recovered $R(t)$ proportions of the population as time $t$ changes.  Under mass action assumptions, an infected individual causes new infections at rate $\hat{\beta} S(t)$, where $\hat{\beta}$ is the per-infected transmission rate and $S$ is the probability the recipient is susceptible.  Recovery to an immune state happens at rate $\gamma$.  The flux of individuals from susceptible to infected to recovered is represented by a flow diagram (figure~\ref{fig:compartments}) making the model conceptually simple.  This leads to a simple mathematical interpretation, the low-dimensional, ordinary differential equation (ODE) system
\[
\dot{S} = -\hat{\beta} I S  \, , \qquad\qquad  \dot{I} = \hat{\beta} I S  - \gamma I  \, , \qquad\qquad  \dot{R} = \gamma I \, .
\]
The dot denotes differentiation in time.  An ODE system allows for easy prediction of details such as early growth rates, final size, and intermediate dynamics.  Using $S+I+R=1$ we reexpress this as
\begin{equation}
\dot{S} = -\hat{\beta} I S  \, , \qquad\qquad  I = 1-S-R  \, , \qquad\qquad  \dot{R} = \gamma I \, ,
\label{eqn:ma}
\end{equation}
The product $IS$ measures the proportion of contacts that are from an infected individual to a susceptible individual.

The MA model often provides a reasonable description of epidemics; however it has well-recognized flaws which cause the frequency of infected to susceptible contacts to vary from $IS$.  We highlight two.  It neglects both \emph{social heterogeneity}, variation in contact rates which can be quite broad~\cite{liljeros,mossong:PLOScontacts}, and \emph{contact duration}, implicitly assuming all contacts are infinitesimally short.  Because of these omissions, model predictions can differ from reality.  For example, due to social heterogeneity, early infections tend to have more contacts~\cite{christakis} and when infected may cause more infections than ``average'' individuals, enhancing the early spread over that predicted by the MA model~\cite{andersson:social,newman:spread,miller:heterogeneity,kenah:second,meyers:sars}.  When contact duration is significant, an infected individual may have already infected its neighbors, reducing its ability to cause new infections.
Because of these assumptions, the MA model predicts the same results for a sexually transmitted disease in a completely monogamous population, a population with serial monogamy, and a population with wide variation in contact levels with mean one.  Intuitively we expect these to produce dramatically different epidemics, but no existing mathematical theory allows analytic comparisons.

Over the past 25 years, attempts have been made to eliminate these assumptions without sacrificing analytical tractability.  With few exceptions (notably~\cite{volz:dynamic_network}) these make an ``all-or-nothing'' assumption about contact duration: contacts are fleeting and never repeated or they never change.  
With fleeting contacts, social heterogeneity is introduced by adding multiple risk groups to the MA model: in extreme cases there are arbitrarily many subgroups (the \emph{Mean Field Social Heterogeneity} model)~\cite{andersonmay,may:hivdynamics, may:dynamics,moreno,pastor-satorras:scale-free}. This model is relatively well understood and can be rigorously reduced to a handful of equations.  With permanent contacts, social heterogeneity is indroduced through static networks~\cite{diekmann:network,andersson:network,newman:spread,kenah:second,miller:heterogeneity}.  Static network results typically give the final size but no dynamic information.  Some attempts to predict dynamics with static networks use Pair Approximation techniques~\cite{eames:pair} relying on approximation of network structures.  More rigorous approaches avoid these approximations, but are more difficult~\cite{volz:cts_time,lindquist,ball:network_eqns,babak:finite}.  Of these, only~\cite{volz:cts_time} yields a closed ODE system, (see also~\cite{miller:volz,house:insights}).  This model lacks an illustration like figure~\ref{fig:compartments}, hampering communication and further of development.  Finite, nonzero contact duration is typically handled through simulation, which is usually too slow to study parameter space.

Although no coherent mathematical structure to study social heterogeneity and contact duration exists, there have been studies collecting this data in real-world contexts~\cite{mossong:PLOScontacts,wallinga:contact_survey,salathe:network,christakis}.   Typically the resulting measurements have been reduced to average contact rates to make the mathematics tractable.  Much of the available and potentially relevant detail is discarded because existing models cannot capture the detail collected.

We find that the appropriate perspective allows us to develop conceptually and mathematically simple models that incorporate social heterogeneity and (arbitrary) contact duration.  This provides a unifying framework for existing models and allows an expanded universe of models.  Our goal in each case is to calculate the susceptible, infected, or recovered proportions of the population, but we find that this can be answered more easily using an equivalent problem.  We ask the question, ``what is the probability that a randomly chosen \emph{test node} $u$ is susceptible, infected, or recovered?''  Because $u$ is chosen randomly, the probability it is susceptible equals the proportion susceptible $S(t)$, and similarly for $I$ and $R$.  If we know $S(t)$, then the initial conditions and $\dot{R} = \gamma I$, \ $I=1-S-R$ determine $I(t)$ and $R(t)$ as in~\eqref{eqn:ma}.

The probability $u$ is susceptible is the probability no neighbor has ever transmitted infection to $u$.  The method to calculate this is the focus of this paper.  This probability depends on how many neighbors $u$ has, the rate its neighbors change, and the probability that a random neighbor is infected at any given time.  Because a random neighbor is likely to have more contacts than a random node, knowing the infected fraction of the population does not give the probability a neighbor of $u$ is infected.  We focus on the probability a random neighbor is infected rather than the probability a random individual is infected.  Once we calculate this, it is straightforward to calculate the probability $u$ is susceptible.  The resulting \emph{edge-based compartmental modeling} approach significantly increases the effects we can study compared to MA models with only a small complexity penalty.

In this paper we consider the spread of epidemics in two general classes of networks, \emph{actual degree networks} (based on \emph{Configuration Model} networks~\cite{MolloyReed,newman:arb_degree,vanderhofstad:randomgraphs}) and \emph{expected degree networks} (based on \emph{Mixed Poisson} [commonly called \emph{Chung-Lu}] networks~\cite{chung:connected,norros:poissonian,britton:generate}).  In both cases we can consider static and dynamic networks.  In actual degree networks, a node is assigned $k$ stubs where $k$ is a random non-negative integer assigned independently for each node from some probability distribution.  Edges are created by pairing stubs from different nodes.  In expected degree networks, a node is assigned $\kappa$ where $\kappa$ is a random non-negative real number.  Edges are assigned between two nodes $u$ and $v$ with probability proportional to $\kappa_u\kappa_v$.  We develop exact differential equations for the large population limit, which we compare with simulations.  Detailed descriptions of the simulation techniques are in the Appendix.  

\begin{table*}
\newcommand\T{\rule{0pt}{2.4ex}}
\newcommand\B{\rule[-1.1ex]{0pt}{0pt}}
\begin{center}
\begin{tabular}{|c|c|c|}
\hline
\T{}\textbf{Model}\B{}   &   \textbf{Population Structure} & \textbf{Section}\\\hline\hline%
\parbox{0.185\columnwidth}{\footnotesize Configuration Model (CM)}&\parbox{0.600\columnwidth}{\footnotesize\T{}Static network with specified degree distribution, assigned using the probability mass function $P(k)$.\B}&\ref{sec:CM}\\\hline
\parbox{0.185\columnwidth}{\footnotesize Dynamic Fixed-Degree (DFD)}&\parbox{0.600\columnwidth}{\footnotesize\T{}Dynamic network for which each node's degree remains a constant value, assigned using $P(k)$.\B}&\ref{sec:DFD}\\\hline%
\parbox{0.185\columnwidth}{\footnotesize Dormant Contact (DC)}&\parbox{0.600\columnwidth}{\footnotesize\T{}Dynamic fixed-degree network incorporating gaps between partnerships; a node may wait before replacing a partner.\B}&\ref{sec:DC}\\\hline %
\parbox{0.185\columnwidth}{\footnotesize\T{}Mixed Poisson model (MP)\B{}}&\parbox{0.600\columnwidth}{\footnotesize\T{}Static network with specified distribution of expected degrees $\kappa$ assigned using the probability density function $\rho(\kappa)$.\B}&\ref{sec:MP}\\\hline
\parbox{0.185\columnwidth}{\footnotesize Dynamic Variable-Degree (DVD)}&\parbox{0.600\columnwidth}{\footnotesize\T{}Dynamic network with degrees varying in time with averages assigned using $\rho(\kappa)$.\B}&\ref{sec:DVD}\\\hline%
\parbox{0.185\columnwidth}{\footnotesize\T{}\raggedright{}Mean Field Social Heterogeneity (MFSH)\B{}} & \parbox{0.600\columnwidth}{\footnotesize\T{}Population with a distribution of contact rates assigned using $P(k)$ or $\rho(\kappa)$ and negligible contact duration.\B}&\ref{sec:ad_MFSH}, \ref{sec:ed_MFSH}\\\hline%
\end{tabular}
\end{center}
\caption{Populations to which we apply edge-based compartmental models.} 
\label{tab:models}
\end{table*}

We summarize the populations we consider in Table~\ref{tab:models}.  We begin by analyzing the simplest edge-based compartmental model in detail, exploring epidemic spread in a static network of known degree distribution, a Configuration Model network.  To derive the equations, we introduce a flow diagram that leads to a simple mathematical formulation.  We next consider disease spreading through dynamic actual degree networks and then static and dynamic expected degree networks.  The template shown here allows us to derive a handful of ODEs for each of these populations.  Unsurprisingly, the stronger our assumptions, the simpler our formulation becomes.  We neglect heterogeneity within the population other than the contact levels, assume the disease has a very simple structure, and assume the population is at equilibrium prior to disease introduction.   The companion papers investigate conditions under which the simpler models are appropriate~\cite{miller:ebcm_hierarchy} and how to apply the technique to more complex population and disease structures~\cite{miller:ebcm_structure}.

\section{Configuration Model Epidemics}
\label{sec:CM}
We demonstrate our approach with Configuration Model (CM) networks.  A CM network is static with a known degree distribution (the distribution of the number of contacts).  We create a CM network with $N$ nodes as follows:  We assign each node $u$ its degree $k_u$ with probability $P(k_u)$ and give it $k_u$ \emph{stubs} (half-edges).  Once all nodes are assigned stubs, we pair stubs randomly into edges.  The probability a randomly selected node $u$ has degree $k$ is $P(k)$. In contrast, the probability a stub of $u$ connects to some stub of $v$ is proportional to $k_v$.  So the probability a randomly selected neighbor of $u$ has degree $k$ is 
\[
P_n(k)=k P(k)/\ave{K}
\]
See~\cite{feld:friends,meyers:sars} for more detail.

We assume the disease transmits from an infected node to a neighbor at rate $\beta$.  If the neighbor is susceptible, it becomes infected.  Infected nodes recover at rate $\gamma$.  Throughout, we assume a large population, small initial proportion infected, the small initial proportion of stubs belonging to infected nodes, and growing outbreak.  Our equations become correct once the number of infections $NI$ has grown large enough to behave deterministically, while the proportion infected $I$ is still small.  While stochastic effects are important, other methods such as branching process approximations~\cite{diekmann:textbook} (which apply in large populations with small numbers infected) maybe more useful.

To calculate $S(t)$, $I(t)$, and $R(t)$ we note that these are the probabilities a random \emph{test node} $u$ is in each state.  We calculate $S(t)$ by noting it is also 
the probability none of $u$'s neighbors has yet transmitted to $u$.   We would like to treat each neighbor as independent, but the probability one neighbor $v$ has become infected is affected by whether another neighbor $w$ of $u$ has transmitted to $u$ since $u$ could infect $v$.  Accounting for this directly requires considerable bookkeeping.  
A simpler approach removes the correlation by assuming $u$ causes no infections.  This does not alter the state of $u$: the probabilities we calculate for $u$ will be the proportion of the population in each state under the original assumption $u$ behaves as any other node, and so this yields an equivalent problem.  Further discussion of this modification is in the Appendix.

\begin{figure}
\parbox{\textwidth}{\parbox{0.5\textwidth}{\scalebox{0.95}{\begin{picture}(0,0)%
\includegraphics{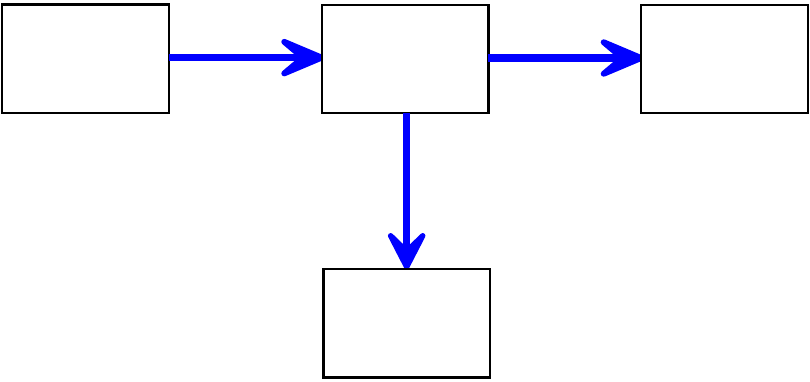}%
\end{picture}%
\setlength{\unitlength}{3947sp}%
\begingroup\makeatletter\ifx\SetFigFontNFSS\undefined%
\gdef\SetFigFontNFSS#1#2#3#4#5{%
  \reset@font\fontsize{#1}{#2pt}%
  \fontfamily{#3}\fontseries{#4}\fontshape{#5}%
  \selectfont}%
\fi\endgroup%
\begin{picture}(3883,1833)(2026,-3544)
\put(2429,-2047){\makebox(0,0)[b]{\smash{{\SetFigFontNFSS{9}{9.6}{\familydefault}{\mddefault}{\updefault}{\color[rgb]{0,0,0}$\phi_S=\frac{\psi'(\theta)}{\psi'(1)}$}%
}}}}
\put(3974,-2047){\makebox(0,0)[b]{\smash{{\SetFigFontNFSS{9}{9.6}{\familydefault}{\mddefault}{\updefault}{\color[rgb]{0,0,0}$\phi_I$}%
}}}}
\put(3974,-3310){\makebox(0,0)[b]{\smash{{\SetFigFontNFSS{9}{9.6}{\familydefault}{\mddefault}{\updefault}{\color[rgb]{0,0,0}$1-\theta$}%
}}}}
\put(5511,-2047){\makebox(0,0)[b]{\smash{{\SetFigFontNFSS{9}{9.6}{\familydefault}{\mddefault}{\updefault}{\color[rgb]{0,0,0}$\phi_R$}%
}}}}
\put(4732,-2224){\makebox(0,0)[b]{\smash{{\SetFigFontNFSS{9}{9.6}{\familydefault}{\mddefault}{\updefault}{\color[rgb]{0,0,0}$\gamma\phi_I$}%
}}}}
\put(4181,-2674){\makebox(0,0)[b]{\smash{{\SetFigFontNFSS{9}{9.6}{\familydefault}{\mddefault}{\updefault}{\color[rgb]{0,0,0}$\beta\phi_I$}%
}}}}
\end{picture}%
}} \hfill
\parbox{0.5\textwidth}{\scalebox{0.95}{\begin{picture}(0,0)%
\includegraphics{standardflux.pdf}%
\end{picture}%
\setlength{\unitlength}{3947sp}%
\begingroup\makeatletter\ifx\SetFigFontNFSS\undefined%
\gdef\SetFigFontNFSS#1#2#3#4#5{%
  \reset@font\fontsize{#1}{#2pt}%
  \fontfamily{#3}\fontseries{#4}\fontshape{#5}%
  \selectfont}%
\fi\endgroup%
\begin{picture}(3883,550)(1651,-1886)
\put(2038,-1700){\makebox(0,0)[b]{\smash{{\SetFigFontNFSS{9}{9.6}{\familydefault}{\mddefault}{\updefault}{\color[rgb]{0,0,0}$S=\psi(\theta)$}%
}}}}
\put(3589,-1700){\makebox(0,0)[b]{\smash{{\SetFigFontNFSS{9}{9.6}{\familydefault}{\mddefault}{\updefault}{\color[rgb]{0,0,0}$I$}%
}}}}
\put(5140,-1700){\makebox(0,0)[b]{\smash{{\SetFigFontNFSS{9}{9.6}{\familydefault}{\mddefault}{\updefault}{\color[rgb]{0,0,0}$R$}%
}}}}
\put(4382,-1509){\makebox(0,0)[b]{\smash{{\SetFigFontNFSS{9}{9.6}{\familydefault}{\mddefault}{\updefault}{\color[rgb]{0,0,0}$\gamma I$}%
}}}}
\end{picture}%
}}} 
\caption{\footnotesize \textbf{Edge-based compartmental modeling for Configuration Model networks.} The flow diagram for a static CM network. (Left) The $\phi_S$, $\phi_I$, and $\phi_R$ compartments represent the probability that a neighbor is susceptible, infected, or recovered and has not transmitted infection. The $1-\theta$ compartment is the probability it has transmitted.  The fluxes between the $\phi$ compartments result from infection or recovery of a neighbor of the test node and the $\phi_I$ to $1-\theta$ flux results from a neighbor transmitting infection to the test node.  (Right) $S$, $I$, and $R$ represent the proportion of the population susceptible, infected, and recovered.  We can find $S$ explicitly, and $I$ and $R$ follow as in the MA model.}
\label{fig:CM}
\end{figure}

We define $\theta(t)$ to be the probability a randomly chosen neighbor has not transmitted to $u$.  Initially $\theta$ is close to $1$.  For large CM networks, neighbors of $u$ are independent. So given its degree $k$, $u$ is susceptible at time $t$ with probability $s(k,\theta(t)) = \theta(t)^k$.  Thus $S(t) = \sum_k P(k) s(k,\theta(t))= \psi(\theta(t))$ where
\[
\psi(x) = \sum_k P(k) x^k
\]
is the probability generating function~\cite{gf} of the degree distribution [the properties of $\psi$ we use are that its derivative is $\sum_k k P(k) x^{k-1}$, its second derivative is $\sum_k k(k-1)P(k)x^{k-2}$, and $\psi'(1)=\ave{K}$].  For many important probability distributions, $\psi$ takes a simple form, which simplifies our examples.  Combining with the flow diagram for $S$, $I$, and $R$ in figure~\ref{fig:CM}, we have
\[
\dot{R} = \gamma I \, ,\qquad\qquad S = \psi(\theta) \, , \qquad\qquad I = 1-S-R
\]
To calculate the new variable $\theta$, 
we break it into three parts; the probability a neighbor $v$ is susceptible at time $t$, $\phi_S$; the probability $v$ is infected at time $t$ but has not transmitted infection to $u$, $\phi_I$; and the probability $v$ has recovered by time $t$ but did not transmit infection to $u$, $\phi_R$.  Then $\theta = \phi_S+\phi_I+\phi_R$.  Initially $\phi_S$ and $\theta$ are approximately $1$ and $\phi_I$, $\phi_R$ are small (they sum to $\theta-\phi_S$). The flow diagram for $\phi_S$, $\phi_I$, $\phi_R$, and $1-\theta$ (figure~\ref{fig:CM}) shows the probability fluxes between these compartments.  
The rate an infected neighbor transmits to $u$ is $\beta$ so the $\phi_I$ to $1-\theta$ flux is $\beta \phi_I$.  We conclude $\dot{\theta} = -\beta \phi_I$.  To find $\phi_I$ we will use $\phi_I = \theta-\phi_S-\phi_R$ and calculate $\phi_S$ and $\phi_R$ explicitly.

The rate an infected neighbor recovers is $\gamma$.  Thus the $\phi_I$ to $\phi_R$ flux is $\gamma\phi_I$.  This is proportional to the flux into $1-\theta$ with the constant of proportionality $\gamma/\beta$.  That is, $\dot{\phi}_R = \gamma \phi_I$, \ $\diff{}{t}(1-\theta) = \beta \phi_I$.
Since $\phi_R$ and $1-\theta$ both begin as approximately $0$, we have $\phi_R=\gamma(1-\theta)/\beta$ in the large population limit.
To find $\phi_S$, recall a neighbor $v$ has degree $k$ with probability $P_n(k)=kP(k)/\ave{K}$.  Given $k$, $v$ is susceptible with probability $\theta^{k-1}$ (we disallow transmission from $u$ so $k-1$ nodes can infect $v$).  A weighted average gives $\phi_S = \sum_kP_n(k)\theta^{k-1}=\sum_k kP(k)\theta^{k-1}/\ave{K} = \psi'(\theta)/\psi'(1)$.  Thus $\phi_I = \theta-\phi_S-\phi_R = \theta -\psi'(\theta)/\psi'(1) - \gamma(1-\theta)/\beta$ and $\dot{\theta}=-\beta\phi_I$ becomes
\[
\dot{\theta} = -\beta\theta + \beta \frac{\psi'(\theta)}{\psi'(1)} + \gamma(1-\theta)
\]
yielding 
\begin{align}
\dot{\theta} &=  -\beta\theta + \beta \frac{\psi'(\theta)}{\psi'(1)} + \gamma (1-\theta)  \, , \\
\dot{R} &= \gamma I  \, , \qquad\qquad  S = \psi(\theta)  \, , \qquad\qquad  I = 1 -S -R \, . 
\end{align}
This captures substantially more population structure than the MA model with only marginally more complexity. 
This is the system of~\cite{miller:volz} and is equivalent to that of~\cite{volz:cts_time}.  It improves on approaches of~\cite{ball:network_eqns,lindquist} which require either $\order(M)$ or $\order(M^2)$ ODEs where $M$ is the (possibly unbounded) maximum degree.  This derivation is simpler than~\cite{volz:cts_time} because we choose variables with a conservation property, simplifying the bookkeeping.

The edge-based compartmental modeling approach we have introduced forms the basis of our paper.  Depending on the network structure, some details will change.  However, we will remain as consistent as possible.

\subsection{$\Ro$ and final size} 
One of the most important parameters for an infectious disease is its basic reproductive number $\Ro$, the average number of infections caused by a node infected early in an epidemic.  When $\Ro<1$ epidemics are impossible, while when $\Ro>1$ they are possible, though not guaranteed.  For this model, we find that $\Ro$ is (see Appendix)
\begin{equation}
\Ro = \frac{\beta}{\beta+\gamma}\frac{\ave{K^2-K}}{\ave{K}} \, .
\end{equation}
We want the expected final size if an epidemic occurs.  We set $\dot{\theta}=0$ and solve
\begin{equation}
\theta(\infty) = \frac{\gamma}{\beta+\gamma} + \frac{\beta}{\beta+\gamma} \frac{\psi'(\theta(\infty))}{\psi'(1)} 
\end{equation}
for $\theta(\infty)$.  If $\Ro>1$ this has two solutions, the larger of which is $\theta=1$ (the pre-disease equilibrium).   We want the smaller solution.   The total fraction of the population infected in the course of an epidemic is $R(\infty)=1-\psi(\theta(\infty))$.  These calculations of $\Ro$ and $R(\infty)$ are in agreement with previous observations~\cite{miller:heterogeneity,newman:spread,andersson:social,andersson:limit,trapman:analytical}.  When $\Ro<1$, our approach breaks down: full details are in the Appendix.

\subsection{Example}
\label{sec:CMex}
\begin{figure}
\parbox{\textwidth}{\includegraphics{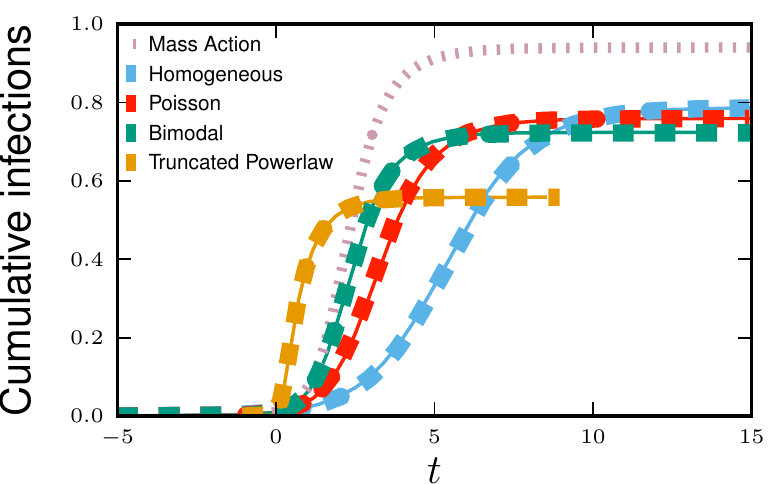} \includegraphics{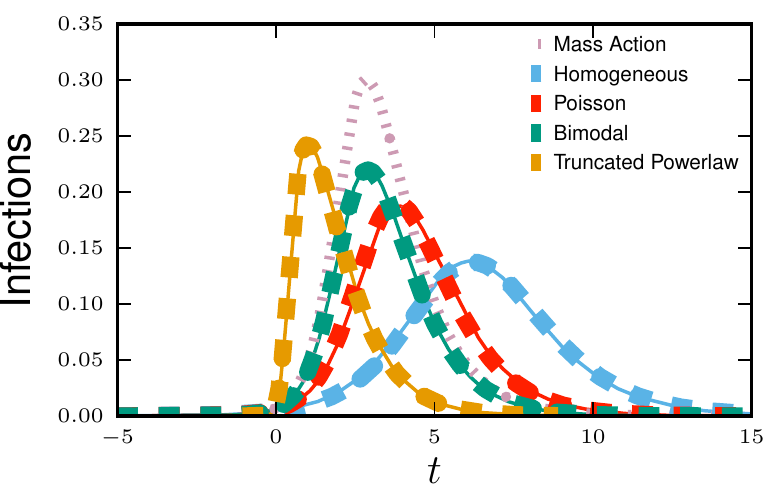}}
\caption{\footnotesize \textbf{Configuration Model Example (section~\ref{sec:CMex}).}  Model predictions (dashed) match simulated epidemics (solid) of the same disease on four Configuration Model networks with $\ave{K} = 5$ and $5\times 10^5$ nodes, but different degree distributions.  Each solid curve is a single simulation.  Time is set so $t=0$ when there is $1$\% cumulative incidence.  The corresponding MA model (short dashes) based on the average degree does not match.}
\label{fig:CMex}
\end{figure}

We consider a disease with $\beta=0.6$ and $\gamma=1$.  Figure~\ref{fig:CMex}  compares simulations with solutions to our ODEs using four different CM networks, each with $5\times 10^5$ nodes and average degree $\ave{K}=5$, but different degree distributions.  In order from latest peak to earliest peak, the networks are: every node has degree $5$, the degree distibution is Poisson with mean $5$, half the nodes have degree $2$ and the other half degree $8$, and finally a truncated powerlaw distribution in which $P(k) \propto k^{-\nu}e^{-k/40}$ where $\nu = 1.418$.    We see that the degree distribution significantly alters the spread, with increased heterogeneity leading to an earlier peak, but generally a smaller epidemic.  Our predictions fit, while the MA model using $\hat{\beta} = \beta\ave{K}$ fails.  

\section{Actual Degree Models}
For the CM networks, each node has a specific number of stubs.  Edges are created by pairing stubs, and no changes are allowed.  In generalizing to other ``actual degree'' models, we assign each node a number of stubs, but allow edges to break and the freed stubs to create new edges.  We consider three limits:  In the first, the Mean Field Social Heterogeneity model, at every moment a stub is connected to a new neighbor.  In the second, the Dynamic Fixed-Degree model, edges last for some time before breaking.  When an edge breaks, the stubs immediately form new edges with stubs from other edges that have just broken.  In the third, the Dormant Contact model, we assume edges break as in the Dynamic Fixed-Degree model, but stubs wait before finding new neighbors.


\subsection{Mean Field Social Heterogeneity}
\label{sec:ad_MFSH}

We analyze the Mean Field Social Heterogeniety (MFSH) model similarly.  We take $\theta$ as the probability a stub has never transmitted infection to the test node $u$ from any neighbor.  To define $\phi_S$, $\phi_I$, and $\phi_R$ we require that the stub has not transmitted infection to $u$ and additionally the current neighbor is susceptible, infected, or recovered.  Since at each moment an individual chooses a new neighbor, the probability of connecting to a node of a given status is the proportion of all stubs belonging to nodes of that status.  We must track the proportion of stubs that belong to susceptible, infected, or recovered nodes $\pi_S$, $\pi_I$, and $\pi_R$.  
Because of the rapid turnover of neighbors, we find that $\phi_S$ is the product of the probability that a stub has not transmitted $\theta$ with the probability it has just joined with a susceptible neighbor $\pi_S$ so $\phi_S = \theta\pi_S$.  Similarly $\phi_I = \theta \pi_I$ and $\phi_R = \theta \pi_R$.  

We create flow diagrams as before.  The $S$, $I$, and $R$ diagram is unchanged, but the diagram for the $\phi$ variables and $1-\theta$ changes.  There are no $\phi_S$ to $\phi_I$ or $\phi_I$ to $\phi_R$ fluxes because of the explicit assumption that the neighbors at any two times are independent.  The change in neighbor status is due to change of neighbor.  The flux into $1-\theta$ from $\phi_I$ is $\beta \phi_I$ as before.  We need a new flow diagram for $\pi_s$, $\pi_I$, and $\pi_R$ similar to that for $S$, $I$, and $R$.  Stubs belonging to infected nodes become stubs belonging to recovered nodes at rate $\gamma$, thus $\dot{\pi}_R = \gamma \pi_I$.   We calculate $\pi_S$ explicitly: the probability a stub belongs to a node of degree $k$ is $kP(k)/\ave{K}$, and the probability the node is susceptible is $\theta^k$.  Taking the weighted average of this we find $\pi_S = \theta \psi'(\theta)/\psi'(1)$.  Finally, $\pi_I = 1-\pi_S-\pi_R$.  

\begin{figure}
\parbox{\textwidth}{\parbox{0.4\textwidth}{\scalebox{0.8}{\begin{picture}(0,0)%
\includegraphics{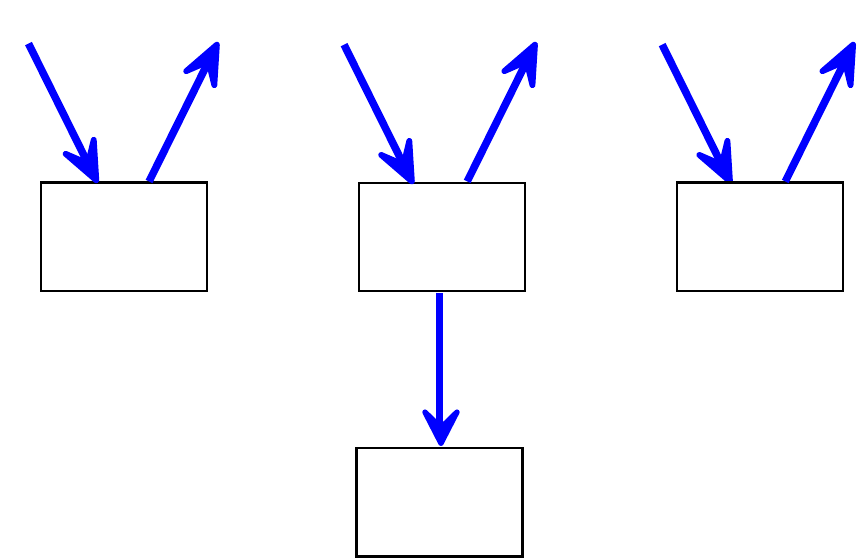}%
\end{picture}%
\setlength{\unitlength}{3947sp}%
\begingroup\makeatletter\ifx\SetFigFontNFSS\undefined%
\gdef\SetFigFontNFSS#1#2#3#4#5{%
  \reset@font\fontsize{#1}{#2pt}%
  \fontfamily{#3}\fontseries{#4}\fontshape{#5}%
  \selectfont}%
\fi\endgroup%
\begin{picture}(3883,2441)(2026,-3544)
\put(2609,-2047){\makebox(0,0)[b]{\smash{{\SetFigFontNFSS{10}{9.6}{\familydefault}{\mddefault}{\updefault}{\color[rgb]{0,0,0}$\phi_S=\theta\pi_S$}%
}}}}
\put(4174,-2047){\makebox(0,0)[b]{\smash{{\SetFigFontNFSS{10}{9.6}{\familydefault}{\mddefault}{\updefault}{\color[rgb]{0,0,0}$\phi_I=\theta\pi_I$}%
}}}}
\put(4074,-3310){\makebox(0,0)[b]{\smash{{\SetFigFontNFSS{10}{9.6}{\familydefault}{\mddefault}{\updefault}{\color[rgb]{0,0,0}$1-\theta$}%
}}}}
\put(5681,-2047){\makebox(0,0)[b]{\smash{{\SetFigFontNFSS{10}{9.6}{\familydefault}{\mddefault}{\updefault}{\color[rgb]{0,0,0}$\phi_R=\theta\pi_R$}%
}}}}
\put(4381,-2674){\makebox(0,0)[b]{\smash{{\SetFigFontNFSS{10}{9.6}{\familydefault}{\mddefault}{\updefault}{\color[rgb]{0,0,0}$\beta\phi_I$}%
}}}}
\end{picture}%
}}\hfill \parbox{0.45\textwidth}{\scalebox{0.9}{\begin{picture}(0,0)%
\includegraphics{standardflux.pdf}%
\end{picture}%
\setlength{\unitlength}{3947sp}%
\begingroup\makeatletter\ifx\SetFigFontNFSS\undefined%
\gdef\SetFigFontNFSS#1#2#3#4#5{%
  \reset@font\fontsize{#1}{#2pt}%
  \fontfamily{#3}\fontseries{#4}\fontshape{#5}%
  \selectfont}%
\fi\endgroup%
\begin{picture}(3883,550)(1651,-1886)
\put(2038,-1700){\makebox(0,0)[b]{\smash{{\SetFigFontNFSS{9}{9.6}{\familydefault}{\mddefault}{\updefault}{\color[rgb]{0,0,0}$S=\psi(\theta)$}%
}}}}
\put(3589,-1700){\makebox(0,0)[b]{\smash{{\SetFigFontNFSS{9}{9.6}{\familydefault}{\mddefault}{\updefault}{\color[rgb]{0,0,0}$I$}%
}}}}
\put(5140,-1700){\makebox(0,0)[b]{\smash{{\SetFigFontNFSS{9}{9.6}{\familydefault}{\mddefault}{\updefault}{\color[rgb]{0,0,0}$R$}%
}}}}
\put(4382,-1509){\makebox(0,0)[b]{\smash{{\SetFigFontNFSS{9}{9.6}{\familydefault}{\mddefault}{\updefault}{\color[rgb]{0,0,0}$\gamma I$}%
}}}}
\end{picture}%
}\\[20pt]
\scalebox{0.9}{\begin{picture}(0,0)%
\includegraphics{standardflux.pdf}%
\end{picture}%
\setlength{\unitlength}{3947sp}%
\begingroup\makeatletter\ifx\SetFigFontNFSS\undefined%
\gdef\SetFigFontNFSS#1#2#3#4#5{%
  \reset@font\fontsize{#1}{#2pt}%
  \fontfamily{#3}\fontseries{#4}\fontshape{#5}%
  \selectfont}%
\fi\endgroup%
\begin{picture}(3883,550)(1651,-1886)
\put(2058,-1700){\makebox(0,0)[b]{\smash{{\SetFigFontNFSS{9}{9.6}{\familydefault}{\mddefault}{\updefault}{\color[rgb]{0,0,0}$\pi_S=\frac{\theta\psi'(\theta)}{\psi'(1)}$}%
}}}}
\put(3589,-1700){\makebox(0,0)[b]{\smash{{\SetFigFontNFSS{9}{9.6}{\familydefault}{\mddefault}{\updefault}{\color[rgb]{0,0,0}$\pi_I$}%
}}}}
\put(5140,-1700){\makebox(0,0)[b]{\smash{{\SetFigFontNFSS{9}{9.6}{\familydefault}{\mddefault}{\updefault}{\color[rgb]{0,0,0}$\pi_R$}%
}}}}
\put(4432,-1509){\makebox(0,0)[b]{\smash{{\SetFigFontNFSS{9}{9.6}{\familydefault}{\mddefault}{\updefault}{\color[rgb]{0,0,0}$\gamma \pi_I$}%
}}}}
\end{picture}%
}}}\\
\caption{\footnotesize \textbf{Mean Field Social Heterogeneity model}.  The flow diagram for the MFSH model (actual degree formulation).  Because contacts are durationless, neighbors do not change status while joined to an individual, so there is no flux between the $\phi$ variables (left).  The new variables $\pi_S$, $\pi_I$, and $\pi_R$ (bottom right) represent the probability that a randomly selected stub belongs to a susceptible, infected, or recovered node.  We can find $\pi_S$ in terms of $\theta$ and then solve for $\pi_I$ and $\pi_R$ in much the same way we solve for $I$ and $R$ in the CM model.  We then find each $\phi$ variable is $\theta$ times the corresponding $\pi$ variable.}
\label{fig:mfsh}
\end{figure}

Combining these observations $\dot{\pi}_R = \gamma \pi_I = \gamma \phi_I/\theta = -(\gamma/\beta) \dot{\theta}/\theta$.  So $\pi_R = -(\gamma/\beta)\ln \theta$ (the constant of integration is $0$).  We have $\pi_I =1-\pi_S-\pi_R= 1-\theta\psi'(\theta)/\psi'(1) + (\gamma/\beta)\ln \theta$ and $\dot{\theta} = -\beta \theta \pi_I$.  Thus 
\begin{align}
\dot{\theta} &= -\beta \theta + \beta \frac{\theta^2\psi'(\theta)}{\psi'(1)} - \theta\gamma \ln \theta\\
\dot{R} &= \gamma I  \, , \qquad\qquad  S = \psi(\theta)  \, , \qquad\qquad  I = 1-S-R
\end{align}

The MFSH model has been considered previously~\cite{andersonmay,may:hivdynamics, may:dynamics,moreno,pastor-satorras:scale-free}, with the population stratified by degree.  Setting $\zeta$ to be the proportion of all stubs which belong to infected nodes (equivalent to $\pi_I$ above), the pre-existing system is
\begin{align*}
 \dot{S}_k &= -\beta kS_k \zeta\\
 \dot{I}_k &= \beta kS_k \zeta - \gamma I_k\\
 \zeta &= \frac{\sum_k kP(k) I_k}{\ave{K}}
 \end{align*}
where $S_k$ and $I_k$ are the probabilities a random individual with $k$ contacts is susceptible or recovered.  A known change of variables reduces this to a few equations equivalent to ours (see Appendix).

\subsubsection{$\Ro$ and final size}
We find 
\begin{equation}
\Ro = \frac{\beta}{\gamma}\frac{\ave{K^2}}{\ave{K}}
\end{equation}
consistent with previous observations~\cite{andersonmay}.  The total proportion infected is $R(\infty) = 1-\psi(\theta(\infty))$ where
\begin{equation}
\theta(\infty) = \exp\left[-\frac{\beta}{\gamma}\left(1-\frac{\theta(\infty)\psi'(\theta(\infty))}{\psi'(1)}\right)\right]
\end{equation}
Full details are in the Appendix.

\subsubsection{Example}
\label{sec:MFSHex}
\begin{figure}
\parbox{\textwidth}{\includegraphics{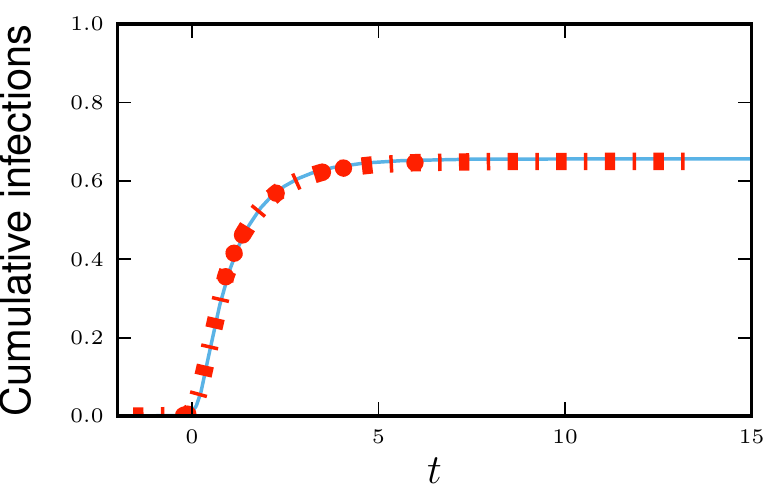}
\includegraphics{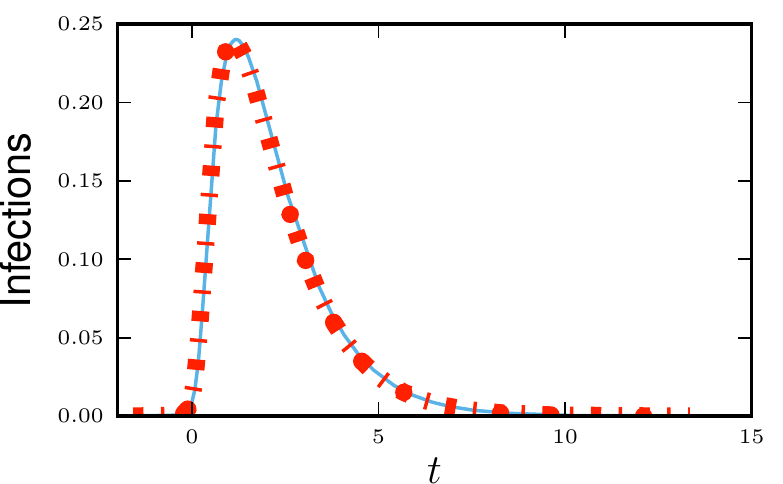}}
\caption{\footnotesize \textbf{Mean Field Social Heterogeneity Example (section~\ref{sec:MFSHex}).}  Model predictions (dashed) match a simulated epidemic (solid) in a population of $5\times 10^5$ nodes.  The solid curve is a single simulation.  Time is set so $t=0$ when there is $1$\% cumulative incidence.}
\label{fig:MFSHex}
\end{figure}

We take a population with degrees $1$, $5$, and $25$.  The proportions are chosen such that an equal number of stubs belong to each class: $P(1)= 25/31$, \ $P(5)=5/31$, and $P(25)=1/31$.  Thus
\[
\psi(x) = \frac{25x+5x^5+x^{25}}{31}
\]
We set $\beta=\gamma=1$ and compare a simulation in a population of $5\times 10^5$ with theory in figure~\ref{fig:MFSHex}.

\subsection{Dynamic Fixed-Degree}
\label{sec:DFD}
The Dynamic Fixed-Degree (DFD) model interpolates between the CM and MFSH models.  We assign each node's degree $k$ as before and pair stubs randomly.  As time progresses, edges break.  The freed stubs immediately join with stubs from other edges that break, a process we refer to as ``edge swapping''.  The rate an edge breaks is $\eta$.  

\begin{figure}
\scalebox{0.9}{\begin{picture}(0,0)%
\includegraphics{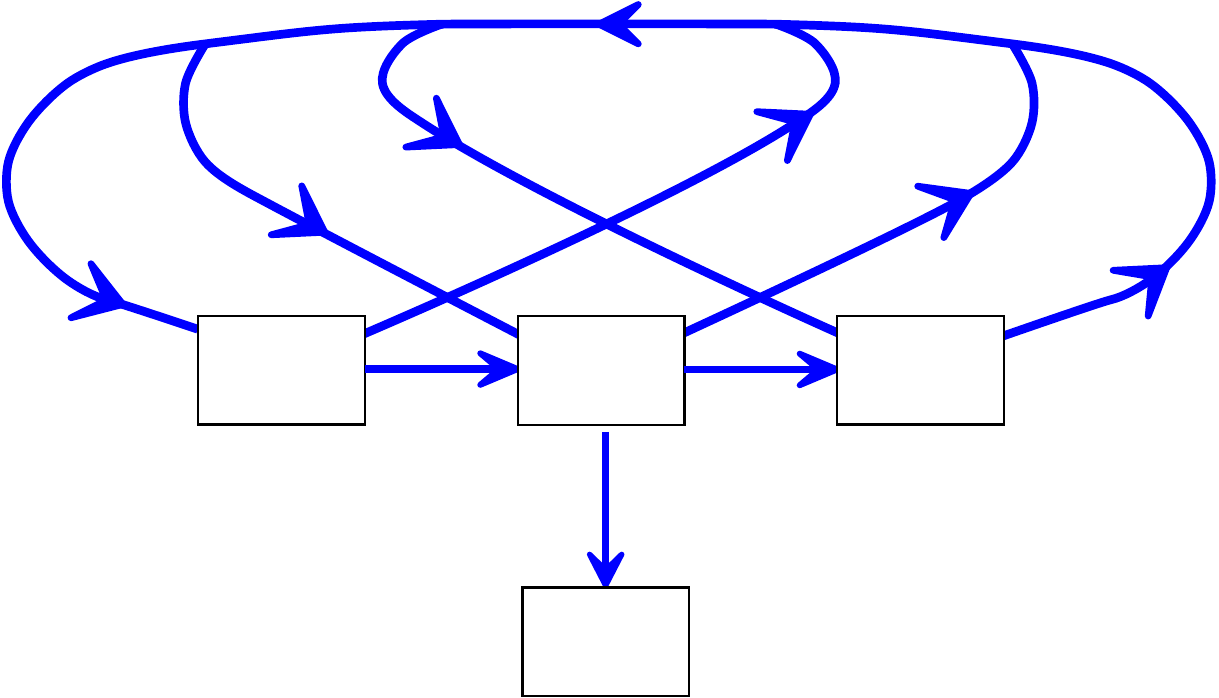}%
\end{picture}%
\setlength{\unitlength}{3947sp}%
\begingroup\makeatletter\ifx\SetFigFontNFSS\undefined%
\gdef\SetFigFontNFSS#1#2#3#4#5{%
  \reset@font\fontsize{#1}{#2pt}%
  \fontfamily{#3}\fontseries{#4}\fontshape{#5}%
  \selectfont}%
\fi\endgroup%
\begin{picture}(5850,3511)(1651,-5494)
\put(4562,-5285){\makebox(0,0)[b]{\smash{{\SetFigFontNFSS{10}{9.6}{\familydefault}{\mddefault}{\updefault}{\color[rgb]{0,0,0}$1-\theta$}%
}}}}
\put(3006,-3980){\makebox(0,0)[b]{\smash{{\SetFigFontNFSS{10}{9.6}{\familydefault}{\mddefault}{\updefault}{\color[rgb]{0,0,0}$\phi_S$}%
}}}}
\put(4576,-3980){\makebox(0,0)[b]{\smash{{\SetFigFontNFSS{10}{9.6}{\familydefault}{\mddefault}{\updefault}{\color[rgb]{0,0,0}$\phi_I$}%
}}}}
\put(6111,-3980){\makebox(0,0)[b]{\smash{{\SetFigFontNFSS{10}{9.6}{\familydefault}{\mddefault}{\updefault}{\color[rgb]{0,0,0}$\phi_R$}%
}}}}
\put(5311,-4120){\makebox(0,0)[b]{\smash{{\SetFigFontNFSS{10}{9.6}{\familydefault}{\mddefault}{\updefault}{\color[rgb]{0,0,0}$\gamma\phi_I$}%
}}}}
\put(4811,-4580){\makebox(0,0)[b]{\smash{{\SetFigFontNFSS{10}{9.6}{\familydefault}{\mddefault}{\updefault}{\color[rgb]{0,0,0}$\beta\phi_I$}%
}}}}
\put(3750,-4340){\makebox(0,0)[t]{\smash{{\SetFigFontNFSS{10}{9.6}{\familydefault}{\mddefault}{\updefault}{\color[rgb]{0,0,0}$\beta\phi_I\phi_S\frac{\psi''(\theta)}{\psi'(\theta)}$}%
}}}}
\put(2085,-3297){\makebox(0,0)[b]{\smash{{\SetFigFontNFSS{10}{9.6}{\familydefault}{\mddefault}{\updefault}{\color[rgb]{0,0,0}$\eta\theta \pi_S$}%
}}}}
\put(3083,-2934){\makebox(0,0)[b]{\smash{{\SetFigFontNFSS{10}{9.6}{\familydefault}{\mddefault}{\updefault}{\color[rgb]{0,0,0}$\eta\theta\pi_I $}%
}}}}
\put(4176,-2824){\makebox(0,0)[b]{\smash{{\SetFigFontNFSS{10}{9.6}{\familydefault}{\mddefault}{\updefault}{\color[rgb]{0,0,0}$\eta\theta\pi_R$}%
}}}}
\put(5704,-2892){\makebox(0,0)[b]{\smash{{\SetFigFontNFSS{10}{9.6}{\familydefault}{\mddefault}{\updefault}{\color[rgb]{0,0,0}$\eta\phi_S$}%
}}}}
\put(4607,-2532){\makebox(0,0)[b]{\smash{{\SetFigFontNFSS{10}{9.6}{\familydefault}{\mddefault}{\updefault}{\color[rgb]{0,0,0}$\eta\theta$}%
}}}}
\put(7433,-3662){\makebox(0,0)[b]{\smash{{\SetFigFontNFSS{10}{9.6}{\familydefault}{\mddefault}{\updefault}{\color[rgb]{0,0,0}$\eta\phi_R$}%
}}}}
\put(6460,-3268){\makebox(0,0)[b]{\smash{{\SetFigFontNFSS{10}{9.6}{\familydefault}{\mddefault}{\updefault}{\color[rgb]{0,0,0}$\eta\phi_I$}%
}}}}
\end{picture}%
} \hfill
\scalebox{0.9}{\begin{picture}(0,0)%
\includegraphics{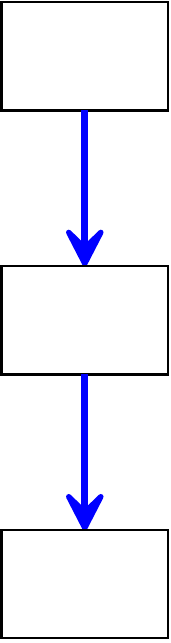}%
\end{picture}%
\setlength{\unitlength}{3947sp}%
\begingroup\makeatletter\ifx\SetFigFontNFSS\undefined%
\gdef\SetFigFontNFSS#1#2#3#4#5{%
  \reset@font\fontsize{#1}{#2pt}%
  \fontfamily{#3}\fontseries{#4}\fontshape{#5}%
  \selectfont}%
\fi\endgroup%
\begin{picture}(817,3083)(2551,-5019)
\put(2946,-2248){\makebox(0,0)[b]{\smash{{\SetFigFontNFSS{9}{9.6}{\familydefault}{\mddefault}{\updefault}{\color[rgb]{0,0,0}$S=\psi(\theta)$}%
}}}}
\put(2946,-3520){\makebox(0,0)[b]{\smash{{\SetFigFontNFSS{9}{9.6}{\familydefault}{\mddefault}{\updefault}{\color[rgb]{0,0,0}$I$}%
}}}}
\put(2946,-4781){\makebox(0,0)[b]{\smash{{\SetFigFontNFSS{9}{9.6}{\familydefault}{\mddefault}{\updefault}{\color[rgb]{0,0,0}$R$}%
}}}}
\put(3137,-4070){\makebox(0,0)[b]{\smash{{\SetFigFontNFSS{9}{9.6}{\familydefault}{\mddefault}{\updefault}{\color[rgb]{0,0,0}$\gamma I$}%
}}}}
\end{picture}%
}\hfill
\scalebox{0.9}{\begin{picture}(0,0)%
\includegraphics{vertical_flux.pdf}%
\end{picture}%
\setlength{\unitlength}{3947sp}%
\begingroup\makeatletter\ifx\SetFigFontNFSS\undefined%
\gdef\SetFigFontNFSS#1#2#3#4#5{%
  \reset@font\fontsize{#1}{#2pt}%
  \fontfamily{#3}\fontseries{#4}\fontshape{#5}%
  \selectfont}%
\fi\endgroup%
\begin{picture}(817,3083)(2551,-5019)
\put(2946,-2248){\makebox(0,0)[b]{\smash{{\SetFigFontNFSS{9}{9.6}{\familydefault}{\mddefault}{\updefault}{\color[rgb]{0,0,0}$\pi_S=\frac{\theta\psi'(\theta)}{\psi'(1)}$}%
}}}}
\put(2946,-3520){\makebox(0,0)[b]{\smash{{\SetFigFontNFSS{9}{9.6}{\familydefault}{\mddefault}{\updefault}{\color[rgb]{0,0,0}$\pi_I$}%
}}}}
\put(2946,-4781){\makebox(0,0)[b]{\smash{{\SetFigFontNFSS{9}{9.6}{\familydefault}{\mddefault}{\updefault}{\color[rgb]{0,0,0}$\pi_R$}%
}}}}
\put(3137,-4070){\makebox(0,0)[b]{\smash{{\SetFigFontNFSS{9}{9.6}{\familydefault}{\mddefault}{\updefault}{\color[rgb]{0,0,0}$\gamma \pi_I$}%
}}}}
\end{picture}%
}\\
\caption{\footnotesize \textbf{Dynamic Fixed-Degree model.}  The flow diagram for the DFD model.  Unlike the CM case, we cannot calculate $\phi_S$ explicitly, so we must calculate the $\phi_S$ to $\phi_I$ flux.}
\label{fig:fd}
\end{figure}

We develop flow diagrams (figure~\ref{fig:fd}) as before.  The $S$, $I$, and $R$ diagram is unchanged.  We again track the probabilities $\pi_S$, $\pi_I$, and $\pi_R$ that a random stub belongs to a susceptible, infected, or recovered node.  The diagram is unchanged.  The diagram for $\theta$ and the $\phi$ variables changes:  We have fluxes from $\phi_S$ to $\phi_I$ and $\phi_I$ to $\phi_R$ representing infection or recovery of the neighbor as in the CM model, but we also have fluxes from $\phi_S$ to $\phi_S$, $\phi_I$, or $\phi_R$ resulting from edge swapping.  We have similar edge swapping fluxes from $\phi_I$ and $\phi_R$.   The flux into $\phi_S$  from edge swapping is $\eta\theta\pi_S$.  The flux out of $\phi_S$  from edge swapping is $\eta \phi_S$.  Similar results hold for $\phi_I$ and $\phi_R$.

Our earlier techniques to find $\phi_I$ break down.  We solve for $\phi_S$ and $\phi_I$ using ODEs.  To complete the system, we need the $\phi_S$ to $\phi_I$ flux.  Consider a neighbor $v$ of our test node $u$ such that: the stub belonging to $u$ never transmitted to $u$ and the stub belonging to $v$ never transmitted to $v$ prior to the $u$-$v$ edge forming
Given this, the probability $v$ is susceptible is $q=\sum_k k P(k)\theta^{k-1}/\ave{K}= \psi'(\theta)/\psi'(1)$.  Thus, given that $v$ is susceptible, $v$ becomes infected at rate
\[
-\frac{\dot{q}}{q} = -\frac{\dot{\theta}\psi''(\theta)/\psi'(1)}{\psi'(\theta)/\psi'(1)}=\beta \phi_I \frac{\psi''(\theta)}{\psi'(\theta)}
\]  
Thus the $\phi_S$ to $\phi_I$ flux is the product of $\phi_S$, the probability a stub has not transmitted infection to the test node and connects to a susceptible node, with  $\beta \phi_I \psi''(\theta)/\psi'(\theta)$, the rate the node becomes infected given that the stub has not transmitted and connects to a susceptible node.  This completes figure~\ref{fig:fd}.

The model requires more equations, but remains relatively simple:
\begin{align}
\dot{\theta} &= -\beta \phi_I \, ,\\
\dot{\phi}_S &= - \beta \phi_I \phi_S\frac{\psi''(\theta)}{\psi'(\theta)}  + \eta\theta\pi_S - \eta \phi_S\, ,  \\
\dot{\phi}_I &= \beta \phi_I \phi_S\frac{\psi''(\theta)}{\psi'(\theta)} + \eta \theta \pi_I - (\beta  + \gamma + \eta) \phi_I \, ,  \\
\dot{\pi}_R &= \gamma \pi_I \, , \qquad\qquad \pi_S = \frac{\theta\psi'(\theta)}{\psi'(1)} \, ,\qquad\qquad 
\pi_I = 1-\pi_R-\pi_S\, ,  \\ 
\dot{R} &= \gamma I \, ,\qquad\qquad S(t) = \psi(\theta) \, ,\qquad\qquad I(t) = 1-S-R \, . 
\end{align}
This is simpler than, but equivalent to, the model of~\cite{volz:dynamic_network}.

\subsubsection{$\Ro$ and final size}
 We find 
 \begin{equation}
 \Ro = \frac{\beta}{(\beta+\eta+\gamma)}\left( \frac{\eta + \gamma}{\gamma} \frac{\ave{K^2-K}}{\ave{K}} + \frac{\eta}{\gamma}\right) \, ,
\end{equation}
We do not find a simple expression for final size.  Instead we must solve the ODEs numerically.  Full details are in the Appendix.

\subsubsection{Example}
\label{sec:FDex}
\begin{figure}
\parbox{\textwidth}{\includegraphics{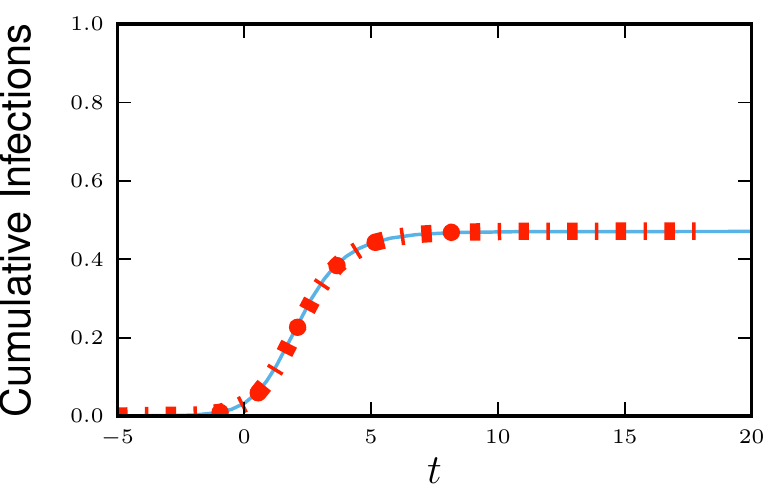}
\includegraphics{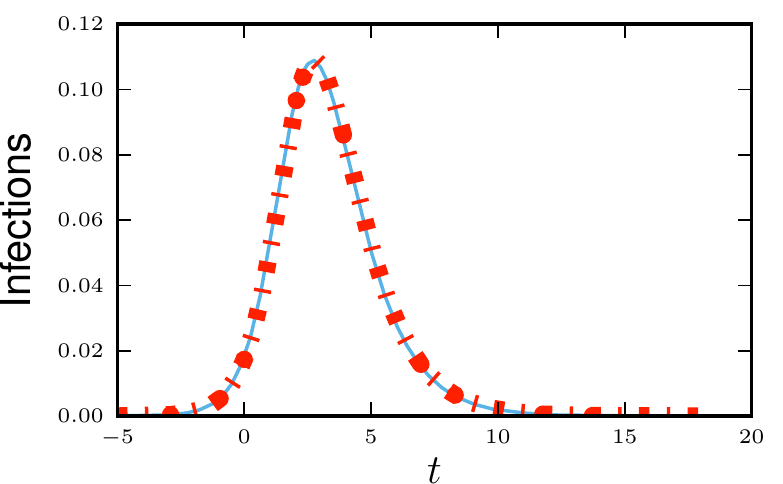}}
\caption{\footnotesize \textbf{Dynamic Fixed-Degree Example (section~\ref{sec:FDex}).}  Model predictions (dashed) match the average of 102 simulated epidemics (solid) in a population of $10^4$ nodes.  For each simulation, time is chosen so that $t=0$ corresponds to $3\%$ cumulative incidence.  Then they are averaged to give the solid curve.}
\label{fig:fdex}
\end{figure}

We choose a population having negative binomial degree distribution $\NB(4,1/3)$ with size $r=4$ and probability $p=1/3$.  Thus $P(k) = \binom{k+r-1}{k}(1-p)^rp^k$.  The mean is $2$ and the variance $3$.  For negative binomial distributions $\psi(x) = [(1-p)/(1-px)]^r$, so
\[
\psi(x) = \left(\frac{2}{3-x}\right)^4
\]
We take $\beta=5/4$, \ $\gamma = 1$, and $\eta=1/2$.  The equations accurately predict the spread (figure~\ref{fig:fdex}).  

Our simulations are slower because we must track edges, so we have used smaller population sizes. To reduce noise, we perform 250 simulations, averaging the 102 that become epidemics.


\subsection{Dormant Contacts}
\label{sec:DC}
We finally generalize the DFD model, allowing stubs to enter a dormant phase after edges break.  This Dormant Contact (DC) model is appropriate for serial monogamy where individuals do not immediately find a new partner.  It is the most general model we present: it reduces to any model of this paper in appropriate limits~\cite{miller:ebcm_hierarchy}.

A node is assigned $k_m$ stubs using the probability mass function $P(k_m)$.  We take $\psi(x) = \sum_{k_m} P(k_m)x^{k_m}$.  A stub is dormant or active depending on whether it is currently connected to a neighbor.  The maximum degree of a node is $k_m$ and the ``active'' and ``dormant'' degrees are $k_a$ and $k_d$ respectively, $k_a+k_d=k_m$.  In addition to $\phi_S$, $\phi_I$, and $\phi_R$ we add $\phi_D$ denoting the probability a stub is dormant and has never transmitted infection from a neighbor, so $\theta=\phi_S+\phi_I+\phi_R+\phi_D$.  Active stubs become dormant at rate $\eta_2$ and dormant stubs become active at rate $\eta_1$.

\begin{figure}
\begin{center}
\scalebox{0.75}{\begin{picture}(0,0)%
\includegraphics{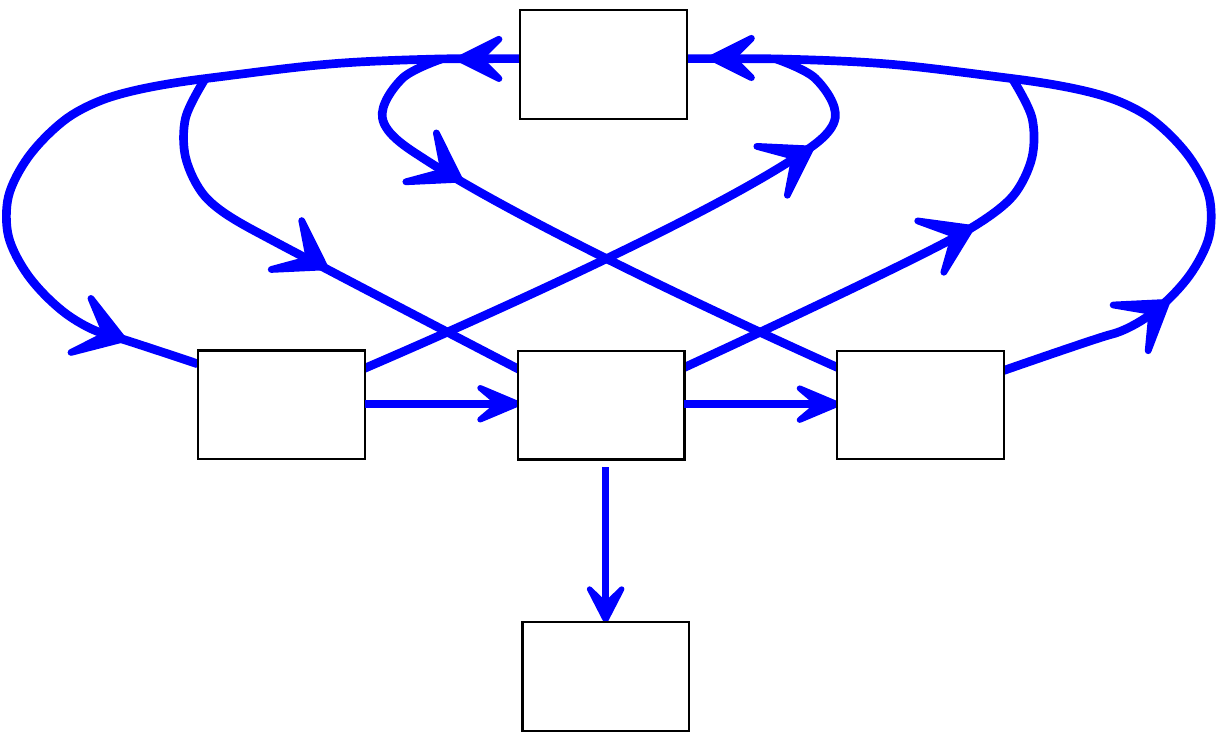}%
\end{picture}%
\setlength{\unitlength}{3947sp}%
\begingroup\makeatletter\ifx\SetFigFontNFSS\undefined%
\gdef\SetFigFontNFSS#1#2#3#4#5{%
  \reset@font\fontsize{#1}{#2pt}%
  \fontfamily{#3}\fontseries{#4}\fontshape{#5}%
  \selectfont}%
\fi\endgroup%
\begin{picture}(5850,3511)(1651,-5494)
\put(4562,-5285){\makebox(0,0)[b]{\smash{{\SetFigFontNFSS{12}{9.6}{\familydefault}{\mddefault}{\updefault}{\color[rgb]{0,0,0}$1-\theta$}%
}}}}
\put(3006,-3980){\makebox(0,0)[b]{\smash{{\SetFigFontNFSS{12}{9.6}{\familydefault}{\mddefault}{\updefault}{\color[rgb]{0,0,0}$\phi_S$}%
}}}}
\put(4576,-3980){\makebox(0,0)[b]{\smash{{\SetFigFontNFSS{12}{9.6}{\familydefault}{\mddefault}{\updefault}{\color[rgb]{0,0,0}$\phi_I$}%
}}}}
\put(6111,-3980){\makebox(0,0)[b]{\smash{{\SetFigFontNFSS{12}{9.6}{\familydefault}{\mddefault}{\updefault}{\color[rgb]{0,0,0}$\phi_R$}%
}}}}

\put(3750,-4380){\makebox(0,0)[t]{\smash{{\SetFigFontNFSS{12}{9.6}{\familydefault}{\mddefault}{\updefault}{\color[rgb]{0,0,0}$\beta\phi_I\phi_S\frac{\psi''(\theta)}{\psi'(\theta)}$}%
}}}}
\put(5311,-4080){\makebox(0,0)[b]{\smash{{\SetFigFontNFSS{12}{9.6}{\familydefault}{\mddefault}{\updefault}{\color[rgb]{0,0,0}$\gamma\phi_I$}%
}}}}
\put(4811,-4580){\makebox(0,0)[b]{\smash{{\SetFigFontNFSS{12}{9.6}{\familydefault}{\mddefault}{\updefault}{\color[rgb]{0,0,0}$\beta\phi_I$}%
}}}}

\put(2155,-3287){\makebox(0,0)[b]{\smash{{\SetFigFontNFSS{12}{9.6}{\familydefault}{\mddefault}{\updefault}{\color[rgb]{0,0,0}$\eta_1 \frac{\pi_S}{\pi} \phi_D$}%
}}}}
\put(3093,-2924){\makebox(0,0)[b]{\smash{{\SetFigFontNFSS{12}{9.6}{\familydefault}{\mddefault}{\updefault}{\color[rgb]{0,0,0}$\eta_1\frac{\pi_I}{\pi} \phi_D$}%
}}}}
\put(4226,-2804){\makebox(0,0)[b]{\smash{{\SetFigFontNFSS{12}{9.6}{\familydefault}{\mddefault}{\updefault}{\color[rgb]{0,0,0}$\eta_1\frac{\pi_R}{\pi}\phi_D$}%
}}}}
\put(4527,-2334){\makebox(0,0)[b]{\smash{{\SetFigFontNFSS{12}{9.6}{\familydefault}{\mddefault}{\updefault}{\color[rgb]{0,0,0}$\phi_D$}%
}}}}
\put(5694,-2892){\makebox(0,0)[b]{\smash{{\SetFigFontNFSS{12}{9.6}{\familydefault}{\mddefault}{\updefault}{\color[rgb]{0,0,0}$\eta_2\phi_S$}%
}}}}
\put(3857,-2082){\makebox(0,0)[b]{\smash{{\SetFigFontNFSS{12}{9.6}{\familydefault}{\mddefault}{\updefault}{\color[rgb]{0,0,0}$\eta_1\phi_D$}%
}}}}
\put(5425,-2082){\makebox(0,0)[b]{\smash{{\SetFigFontNFSS{12}{9.6}{\familydefault}{\mddefault}{\updefault}{\color[rgb]{0,0,0}$\eta_2(\theta-\phi_D)$}%
}}}}
\put(7383,-3682){\makebox(0,0)[b]{\smash{{\SetFigFontNFSS{12}{9.6}{\familydefault}{\mddefault}{\updefault}{\color[rgb]{0,0,0}$\eta_2\phi_R$}%
}}}}
\put(6480,-3288){\makebox(0,0)[b]{\smash{{\SetFigFontNFSS{12}{9.6}{\familydefault}{\mddefault}{\updefault}{\color[rgb]{0,0,0}$\eta_2\phi_I$}%
}}}}
\end{picture}%
} \hfill \scalebox{0.75}{\begin{picture}(0,0)%
\includegraphics{vertical_flux.pdf}%
\end{picture}%
\setlength{\unitlength}{3947sp}%
\begingroup\makeatletter\ifx\SetFigFontNFSS\undefined%
\gdef\SetFigFontNFSS#1#2#3#4#5{%
  \reset@font\fontsize{#1}{#2pt}%
  \fontfamily{#3}\fontseries{#4}\fontshape{#5}%
  \selectfont}%
\fi\endgroup%
\begin{picture}(817,3083)(2551,-5019)
\put(2946,-2248){\makebox(0,0)[b]{\smash{{\SetFigFontNFSS{10}{9.6}{\familydefault}{\mddefault}{\updefault}{\color[rgb]{0,0,0}$S=\psi(\theta)$}%
}}}}
\put(2946,-3520){\makebox(0,0)[b]{\smash{{\SetFigFontNFSS{10}{9.6}{\familydefault}{\mddefault}{\updefault}{\color[rgb]{0,0,0}$I$}%
}}}}
\put(2946,-4781){\makebox(0,0)[b]{\smash{{\SetFigFontNFSS{10}{9.6}{\familydefault}{\mddefault}{\updefault}{\color[rgb]{0,0,0}$R$}%
}}}}
\put(3137,-4070){\makebox(0,0)[b]{\smash{{\SetFigFontNFSS{10}{9.6}{\familydefault}{\mddefault}{\updefault}{\color[rgb]{0,0,0}$\gamma I$}%
}}}}
\end{picture}%
}\hfill
\scalebox{0.75}{\begin{picture}(0,0)%
\includegraphics{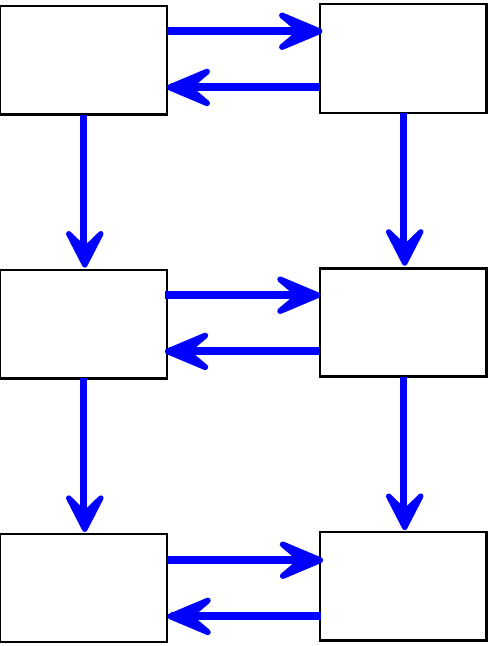}%
\end{picture}%
\setlength{\unitlength}{3947sp}%
\begingroup\makeatletter\ifx\SetFigFontNFSS\undefined%
\gdef\SetFigFontNFSS#1#2#3#4#5{%
  \reset@font\fontsize{#1}{#2pt}%
  \fontfamily{#3}\fontseries{#4}\fontshape{#5}%
  \selectfont}%
\fi\endgroup%
\begin{picture}(2350,3344)(1426,-4809)
\put(1485,-2072){\makebox(0,0)[b]{\smash{{\SetFigFontNFSS{12}{9.6}{\familydefault}{\mddefault}{\updefault}{\color[rgb]{0,0,0}\parbox{0.02\textwidth}{~~$\xi_S=\frac{(\theta-\phi_D)\psi'(\theta)}{\psi'(1)}$}}%
}}}}
\put(1821,-3273){\makebox(0,0)[b]{\smash{{\SetFigFontNFSS{12}{9.6}{\familydefault}{\mddefault}{\updefault}{\color[rgb]{0,0,0}$\xi_I$}%
}}}}
\put(1821,-4540){\makebox(0,0)[b]{\smash{{\SetFigFontNFSS{12}{9.6}{\familydefault}{\mddefault}{\updefault}{\color[rgb]{0,0,0}$\xi_R$}%
}}}}
\put(3366,-4540){\makebox(0,0)[b]{\smash{{\SetFigFontNFSS{12}{9.6}{\familydefault}{\mddefault}{\updefault}{\color[rgb]{0,0,0}$\pi_R$}%
}}}}
\put(3360,-3273){\makebox(0,0)[b]{\smash{{\SetFigFontNFSS{12}{9.6}{\familydefault}{\mddefault}{\updefault}{\color[rgb]{0,0,0}$\pi_I$}%
}}}}
\put(3165,-2072){\makebox(0,0)[b]{\smash{{\SetFigFontNFSS{12}{9.6}{\familydefault}{\mddefault}{\updefault}{\color[rgb]{0,0,0}\parbox{0.02\textwidth}{~~$\pi_S=\frac{\phi_D\psi'(\theta)}{\psi'(1)}$}}%
}}}}
\put(1994,-3896){\makebox(0,0)[b]{\smash{{\SetFigFontNFSS{12}{9.6}{\familydefault}{\mddefault}{\updefault}{\color[rgb]{0,0,0}$\gamma\xi_I$}%
}}}}
\put(3544,-3896){\makebox(0,0)[b]{\smash{{\SetFigFontNFSS{12}{9.6}{\familydefault}{\mddefault}{\updefault}{\color[rgb]{0,0,0}$\gamma\pi_I$}%
}}}}
\put(2533,-4237){\makebox(0,0)[b]{\smash{{\SetFigFontNFSS{12}{9.6}{\familydefault}{\mddefault}{\updefault}{\color[rgb]{0,0,0}$\eta_2\xi_R$}%
}}}}
\put(2533,-2954){\makebox(0,0)[b]{\smash{{\SetFigFontNFSS{12}{9.6}{\familydefault}{\mddefault}{\updefault}{\color[rgb]{0,0,0}$\eta_2\xi_I$}%
}}}}
\put(2533,-1685){\makebox(0,0)[b]{\smash{{\SetFigFontNFSS{12}{9.6}{\familydefault}{\mddefault}{\updefault}{\color[rgb]{0,0,0}$\eta_2\xi_S$}%
}}}}
\put(2533,-2303){\makebox(0,0)[b]{\smash{{\SetFigFontNFSS{12}{9.6}{\familydefault}{\mddefault}{\updefault}{\color[rgb]{0,0,0}$\eta_1\pi_S$}%
}}}}
\put(2533,-3597){\makebox(0,0)[b]{\smash{{\SetFigFontNFSS{12}{9.6}{\familydefault}{\mddefault}{\updefault}{\color[rgb]{0,0,0}$\eta_1\pi_I$}%
}}}}
\put(2533,-4864){\makebox(0,0)[b]{\smash{{\SetFigFontNFSS{12}{9.6}{\familydefault}{\mddefault}{\updefault}{\color[rgb]{0,0,0}$\eta_1\pi_R$}%
}}}}
\end{picture}%
} 
\end{center}
\caption{\footnotesize\textbf{Dormant Contact model.}  A flow diagram accounting for the dormant stage.  (Left) Movement of stubs between different stages, including dormant.  Stubs are classified by whether they have received infection and the status (or existence) of the current neighbor. (Middle) The flow of individuals between different states.  (Right)  Movement of stubs between states with stubs classified based on the status of the node they belong to.}
\label{fig:dormant}
\end{figure}

We now develop flow diagrams (figure~\ref{fig:dormant}).  The diagram for $S$, $I$, and $R$ is as before.  The diagram for $\theta$ and the $\phi$ variables is similar to the DFD model, but with the new compartment $\phi_D$.  The fluxes associated with edge breaking are at rate $\eta_2$ times $\phi_S$, $\phi_I$, or $\phi_R$ and go from the appropriate compartment into $\phi_D$, for a total of $\eta_2(\theta-\phi_D)$.  
To describe fluxes due to edge creation, we generalize the definitions of $\pi_S$, $\pi_I$, and $\pi_R$ to give the probability a stub is dormant (and thus available to form a new contact) and belongs to a susceptible, infected, or recovered node, with $\pi = \pi_S+\pi_I+\pi_R$ the probability a stub is dormant.  The probability a new neighbor is susceptible, infected, or recovered is $\pi_S/\pi$, $\pi_I/\pi$, and $\pi_R/\pi$ respectively.  The fluxes associated with edge creation occur at total rate $\eta_1\phi_D$, with proportions $\pi_S/\pi$, $\pi_I/\pi$, and $\pi_R/\pi$ into $\phi_S$, $\phi_I$, and $\phi_R$ respectively.

The flow diagram for the $\pi$ variables is related to that for the DFD model, but we must account for active and dormant stubs.  We use $\xi_S$, $\xi_I$, and $\xi_R$ to be the probabilities a stub is active and belongs to each type of node, with $1-\pi=\xi = \xi_S + \xi_I + \xi_R$ the probability a stub is active.  The $\pi_S$ to $\xi_S$ and $\xi_S$ to $\pi_S$ fluxes are $\eta_1\pi_S$ and $\eta_2\xi_S$ respectively.  Similar results hold for the other compartments.  We can use this to show $\dot{\xi}=-\eta_2\xi+\eta_1\pi=-\eta_2\xi+\eta_1(1-\xi)$ from which we can conclude that (at equilibrium) $\pi = \eta_2/(\eta_1+\eta_2)$ and $\xi = \eta_1/(\eta_1+\eta_2)$.  The fluxes from $\xi_I$ and $\pi_I$ to $\xi_R$ and $\pi_R$ respectively represent recovery of the node the stub belongs to and so are $\gamma\xi_I$ and $\gamma \pi_I$ respectively. 

We can calculate $\xi_S$ and $\pi_S$ explicitly. The probability a dormant stub belongs to a susceptible node is $\pi_S=\phi_D\psi'(\theta)/\psi'(1)$ where $\phi_D$ is the probability the dormant stub has never received infection, and $\psi'(\theta)/\psi'(1)$ is the probability that none of the other stubs have received infection.  Similarly, the probability an active stub belongs to a susceptible node is $\xi_S= (\theta-\phi_D)\psi'(\theta)/\psi'(1)$.     We can use $\pi_I = \pi - \pi_S-\pi_R$ and $\xi_I = \xi - \xi_S-\xi_R$ to simplify the system further.

Our new equations are
\begin{align}
\dot{\theta} &= -\beta \phi_I\, , \\
\dot{\phi}_S &= - \beta \phi_I \phi_S \frac{\psi''(\theta)}{\psi'(\theta)} + \eta_1 \frac{\pi_S}{\pi}\phi_D - \eta_2\phi_S\, , \\
\dot{\phi}_I &= \beta \phi_I\phi_S\frac{\psi''(\theta)}{\psi'(\theta)} + \eta_1 \frac{\pi_I}{\pi} \phi_D - (\eta_2+\beta+\gamma)\phi_I \, , \\
\dot{\phi}_D &= \eta_2(\theta-\phi_D) - \eta_1 \phi_D\, , \\
\dot{\xi}_R &= -\eta_2 \xi_R + \eta_1 \pi_R +\gamma \xi_I \, , \qquad \xi_S = (\theta-\phi_D)\frac{\psi'(\theta)}{\psi'(1)} \, , \qquad \xi_I = \xi- \xi_S - \xi_R\, , \\
\dot{\pi}_R &= \eta_2 \xi_R-\eta_1\pi_R + \gamma \pi_I\, , \qquad \pi_S = \phi_D \frac{\psi'(\theta)}{\psi'(1)}  \, , \qquad \pi_I = \pi - \pi_S - \pi_R\, , \\
\xi &= \frac{\eta_1}{\eta_1+\eta_2} \, , \qquad \pi = \frac{\eta_2}{\eta_1+\eta_2}\, , \\
\dot{R} &= \gamma I \, , \qquad\qquad S = \psi(\theta) \, , \qquad\qquad I = 1-S-R \, .
\end{align}

\subsubsection{$\Ro$ and final size}
We can show that 
\begin{equation}
\Ro = 
\frac{\beta}{\beta+\eta_2+\gamma}\left(\frac{\ave{K_m^2-K_m}}{\ave{K_m}} \frac{\eta_1}{\eta_1+\eta_2}\frac{\eta_2+\gamma}{\gamma} + \frac{\eta_1\eta_2}{\gamma(\gamma+\eta_1+\eta_2)}\right)
\end{equation}
However, we have not found a simple final size relation.  Details are in the Appendix.
\subsubsection{Example}
\label{sec:DCex}
\begin{figure}
\parbox{0.48\textwidth}{\includegraphics[width=0.48\textwidth]{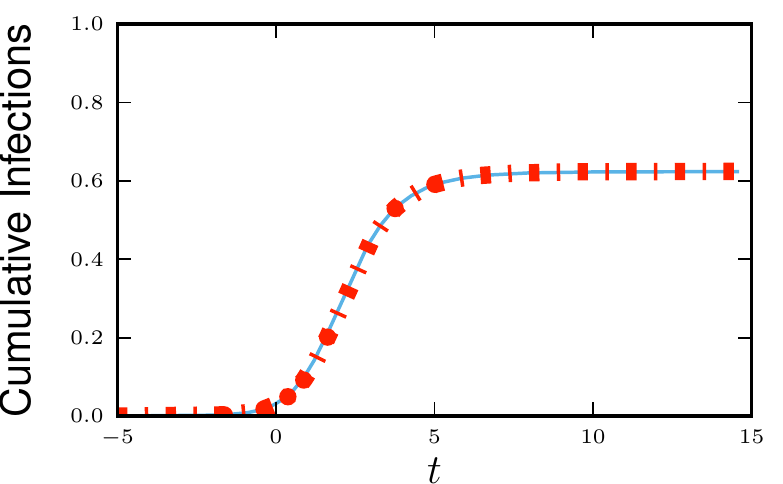}}\hfill
\parbox{0.48\textwidth}{\includegraphics[width=0.48\textwidth]{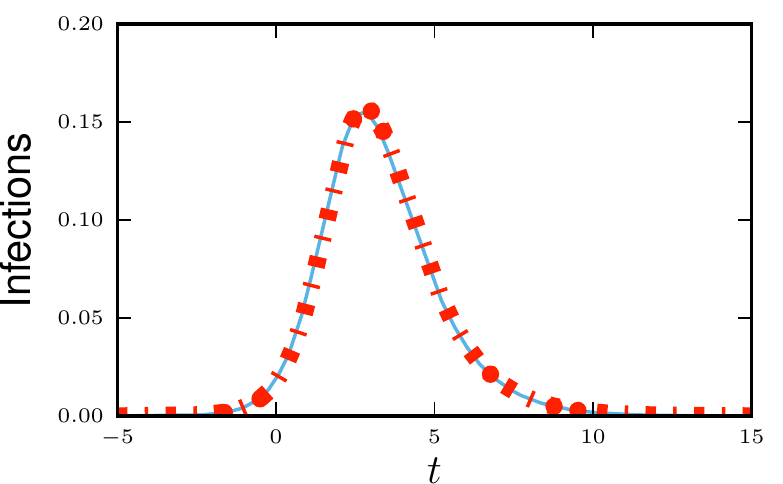}}
\caption{\footnotesize\textbf{Dormant Contact example (section~\ref{sec:DCex}).} The average of 155 simulated epidemics in a population of 5000 nodes (solid) with a Poisson maximum degree distribution of mean 3 compared with theory (dashed).  Simulations are shifted in time so that $t=0$ corresponds to $3\%$ cumulative incidence and then averaged.     Because stochastic effects are not negligible, the individual peaks are not perfectly aligned, so the averaging has a small, but noticeable, effect to reduce and broaden the simulated peak.  In larger populations this disappears.} 
\label{fig:deex}
\end{figure}

In figure~\ref{fig:deex}, we consider the spread of a disease through a network with dormant edges.  The distribution of $k_m$ is Poisson with mean $3$
\[
\psi(x) = e^{-3(1-x)}
\]
  The parameters are $\beta = 2$, \ $\gamma = 1$, \ $\eta_1 = \gamma$, and $\eta_2 = \gamma/2$.  Simulations are again slow, so we use a smaller population with 322 simulations, and average the 155 that became epidemics.

\section{Expected Degree Models}
We now consider SIR diseases spreading through ``expected degree'' networks.  In these networks, each individual has an expected degree $\kappa$, which need not be an integer.  It is assigned using the probability density function $\rho(\kappa)$.  Edges are placed between two nodes with probability proportional to the expected degrees of the two nodes.  In the actual degree models, once a stub belonging to $u$ was joined into an edge, it became unavailable for other edges: the existence of a  $u$-$v$ edge reduced the ability of $u$ to form other edges.  In  contrast for expected degree models, edges are assigned independently: a $u$-$v$ edge does not affect whether $u$ forms other edges.  Similarly to before, a neighbor will tend to have larger expected degree than a randomly chosen edge.  The probability density function for a neighbor to have expected degree $\kappa$ is $\rho_n(\kappa) = \kappa \rho(\kappa)/\ave{K}$.  

Our approach remains similar.  We consider a randomly chosen test node $u$ which cannot infect its neighbors, and calculate the probability $u$ is susceptible.  
We first consider the spread of disease through static ``Mixed Poisson'' networks (also called Chung-Lu networks), for which an edge from $u$ to $v$ exists with probability $\kappa_u\kappa_v/(N-1)\ave{K}$.   We then consider the expected degree formulation of Mean Field Social Heterogeneity.  Following this, we consider the more general Dynamic Variable-Degree model, for which a node creates and deletes edges as independent events, unlike the DFD model where deleted edges were instantantly replaced.  The Mixed Poisson model produces equations effectively identical to the CM equations.  The Mean Field Social heterogeneity equations differ somewhat from the actual degree version, but may be shown~\cite{miller:ebcm_hierarchy} to be formally equivalent.  The Dynamic Variable-Degree equations are simpler than the DFD equations, and it may be more realistic because it does not enforce constant degree for an individual.

\subsection{Mixed Poisson}
\label{sec:MP}
We now consider the Mixed Poisson (MP) model.  In this model, each node is assigned an expected degree $\kappa$ using the probability density function $\rho(\kappa)$.  A $u$-$v$ edge exists with probability $\kappa_u\kappa_v/(N-1)\ave{K}$ independently of other edges.  We use the name ``Mixed Poisson'' because at karge $N$ the actual degree of nodes with expected degree $\kappa$ is chosen from a Poisson distribution with mean $\kappa$.  The degree distribution is a mixture of Poisson distributions.

\begin{figure}
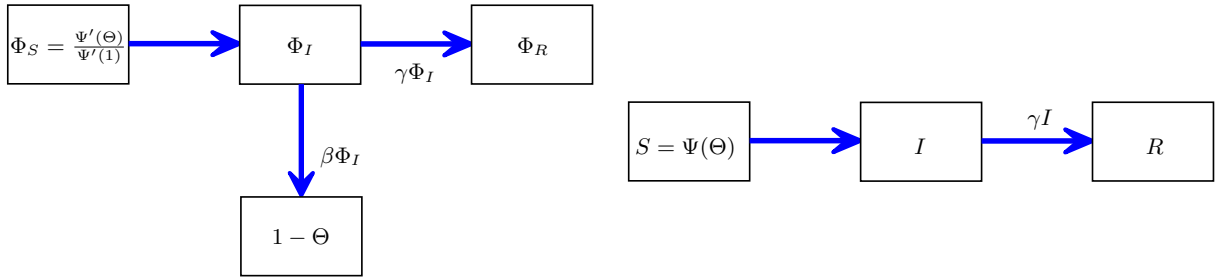

\parbox{\textwidth}{\parbox{0.5\textwidth}{\scalebox{0.95}{\begin{picture}(0,0)%
\includegraphics{edgeflux.pdf}%
\end{picture}%
\setlength{\unitlength}{3947sp}%
\begingroup\makeatletter\ifx\SetFigFontNFSS\undefined%
\gdef\SetFigFontNFSS#1#2#3#4#5{%
  \reset@font\fontsize{#1}{#2pt}%
  \fontfamily{#3}\fontseries{#4}\fontshape{#5}%
  \selectfont}%
\fi\endgroup%
\begin{picture}(3883,1833)(2026,-3544)
\put(2439,-2047){\makebox(0,0)[b]{\smash{{\SetFigFontNFSS{9}{9.6}{\familydefault}{\mddefault}{\updefault}{\color[rgb]{0,0,0}$\Phi_S=\frac{\Psi'(\Theta)}{\Psi'(1)}$}%
}}}}
\put(3974,-2047){\makebox(0,0)[b]{\smash{{\SetFigFontNFSS{9}{9.6}{\familydefault}{\mddefault}{\updefault}{\color[rgb]{0,0,0}$\Phi_I$}%
}}}}
\put(3974,-3310){\makebox(0,0)[b]{\smash{{\SetFigFontNFSS{9}{9.6}{\familydefault}{\mddefault}{\updefault}{\color[rgb]{0,0,0}$1-\Theta$}%
}}}}
\put(5511,-2047){\makebox(0,0)[b]{\smash{{\SetFigFontNFSS{9}{9.6}{\familydefault}{\mddefault}{\updefault}{\color[rgb]{0,0,0}$\Phi_R$}%
}}}}
\put(4732,-2224){\makebox(0,0)[b]{\smash{{\SetFigFontNFSS{9}{9.6}{\familydefault}{\mddefault}{\updefault}{\color[rgb]{0,0,0}$\gamma\Phi_I$}%
}}}}
\put(4241,-2774){\makebox(0,0)[b]{\smash{{\SetFigFontNFSS{9}{9.6}{\familydefault}{\mddefault}{\updefault}{\color[rgb]{0,0,0}$\beta\Phi_I$}%
}}}}
\end{picture}%
}}%
\parbox{0.5\textwidth}{\scalebox{0.95}{\begin{picture}(0,0)%
\includegraphics{standardflux.pdf}%
\end{picture}%
\setlength{\unitlength}{3947sp}%
\begingroup\makeatletter\ifx\SetFigFontNFSS\undefined%
\gdef\SetFigFontNFSS#1#2#3#4#5{%
  \reset@font\fontsize{#1}{#2pt}%
  \fontfamily{#3}\fontseries{#4}\fontshape{#5}%
  \selectfont}%
\fi\endgroup%
\begin{picture}(3883,550)(1651,-1886)
\put(2038,-1700){\makebox(0,0)[b]{\smash{{\SetFigFontNFSS{9}{9.6}{\familydefault}{\mddefault}{\updefault}{\color[rgb]{0,0,0}$S=\Psi(\Theta)$}%
}}}}
\put(3589,-1700){\makebox(0,0)[b]{\smash{{\SetFigFontNFSS{9}{9.6}{\familydefault}{\mddefault}{\updefault}{\color[rgb]{0,0,0}$I$}%
}}}}
\put(5140,-1700){\makebox(0,0)[b]{\smash{{\SetFigFontNFSS{9}{9.6}{\familydefault}{\mddefault}{\updefault}{\color[rgb]{0,0,0}$R$}%
}}}}
\put(4382,-1509){\makebox(0,0)[b]{\smash{{\SetFigFontNFSS{9}{9.6}{\familydefault}{\mddefault}{\updefault}{\color[rgb]{0,0,0}$\gamma I$}%
}}}}
\end{picture}%
}}}
\caption{\footnotesize\textbf{Mixed Poisson model.} (Left) The flux of edges for a static Mixed Poisson network. (Right) The flux of individuals through the different stages}
\label{fig:MP}
\end{figure}

Consider two nodes $u$ and $v$ whose expected degrees are $\kappa_u$ and $\kappa_v=\kappa_u + \Delta \kappa$ with $\Delta \kappa \ll 1$.  Our question is, how much does the additional $\Delta \kappa$ reduce the probability $v$ is susceptible?  At leading order it contributes an extra edge to $v$ with probability $\Delta \kappa$, and we may assume it contributes at most one additional edge.  We define $\Theta$ to be the probability an edge has not transmitted infection.  With probability $\Theta \Delta \kappa$ there is an additional edge which has not transmitted.  The probability the extra $\Delta \kappa$ either does not contribute an edge or contributes an edge which has not transmitted is $1-\Delta \kappa + \Theta \Delta \kappa = 1-(1-\Theta)\Delta \kappa$.  If $s(\kappa,t)$ is the probability a node of expected degree $\kappa$ is susceptible, then we have $s(\kappa+\Delta\kappa,t) = s(\kappa,t) [1-(1-\Theta)\Delta \kappa]$.  Taking $\Delta \kappa \to 0$, this becomes
$\partial s/\partial\kappa = -(1-\Theta)s$.  Thus $s(\kappa,t) = \exp[-\kappa(1-\Theta)]$ and $S(t) = \Psi(\Theta(t))$ where
\[
\Psi(x) = \int_0^\infty e^{-\kappa(1-x)} \rho(\kappa) \, \mathrm{d}\kappa
\]
Note that this is the Laplace transform of $\rho$ evaluated at $1-x$.  As before, figure~\ref{fig:MP} gives
\[
\dot{R} = \gamma I \, , \qquad\qquad S=\Psi(\Theta) \, , \qquad\qquad I = 1-S-R
\]
and we need $\Theta(t)$.

We follow the CM approach almost exactly.  The value of $\Theta$ is the probability an edge has not transmitted to the test node $u$.  We define $\Phi_S$, $\Phi_I$, and $\Phi_R$ to be the probabilities an edge has not transmitted to $u$ and connects to either a susceptible, infected, or recovered node, so $\Theta = \Phi_S+\Phi_I+\Phi_R$.  To calculate $\Phi_S$, we observe that a neighbor $v$ of $u$ with expected degree $\kappa$ has the same probability of having an edge to any $w \neq u$ as any other node of expected degree $\kappa$ because edges are created independently of one another.  Thus given $\kappa$, $v$ is susceptible with probability $s(\kappa,t)$.  Taking the weighted average over all $\kappa$ gives
$\Phi_S = \int_0^\infty \rho_n(\kappa) s(\kappa,t) \, \mathrm{d} t = \int_0^\infty \kappa \exp[-\kappa(1-\Theta)] \rho(\kappa)\, \mathrm{d}\kappa / \ave{K} = \Psi'(\Theta)/\Psi'(1)$.  The same techniques as for the CM networks give $\Phi_R = \gamma (1-\Theta)/\beta$.  Our equations are
\begin{align*}
\dot{\Theta} &=  -\beta\Theta + \beta \frac{\Psi'(\Theta)}{\Psi'(1)} + \gamma (1-\Theta)  \, , \\
\dot{R} &= \gamma I  \, , \qquad\qquad  S = \Psi(\Theta)  \, , \qquad\qquad  I = 1 -S -R \, . 
\end{align*}
These equations are almost identical to those of CM epidemics except that $\Psi$ and $\Theta$ replace $\psi$ and $\theta$ .  This is not coincidence.  In fact the MP networks are a special case of CM networks~\cite{miller:ebcm_hierarchy}.

\subsubsection{$\Ro$ and final size} 
We find
\[
\Ro = \frac{\beta}{\beta+\gamma}\frac{\ave{\hat{K}^2}}{\ave{K}} \, ,
\] 
where $\ave{\hat{K}^2}$ is the average of $\kappa^2$ (which equals the average of $k^2-k$) and $\ave{K}$ is the average of $\kappa$ (which equals the average of $k$).  The total proportion infected by an epidemic is 
$R(\infty) = 1-\Psi(\Theta(\infty))$, where
\[
\Theta(\infty) = \frac{\gamma}{\beta+\gamma} + \frac{\beta}{\beta+\gamma} \frac{\Psi'(\Theta(\infty))}{\Psi'(1)}  \, .
\]
Full details are in the Appendix.

\subsubsection{Example}
\label{sec:MPex}
\begin{figure}
\parbox{\textwidth}{\includegraphics{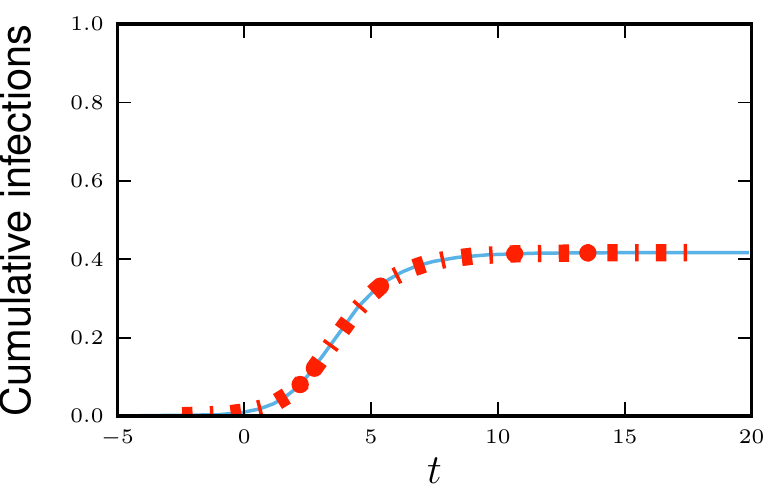}
\includegraphics{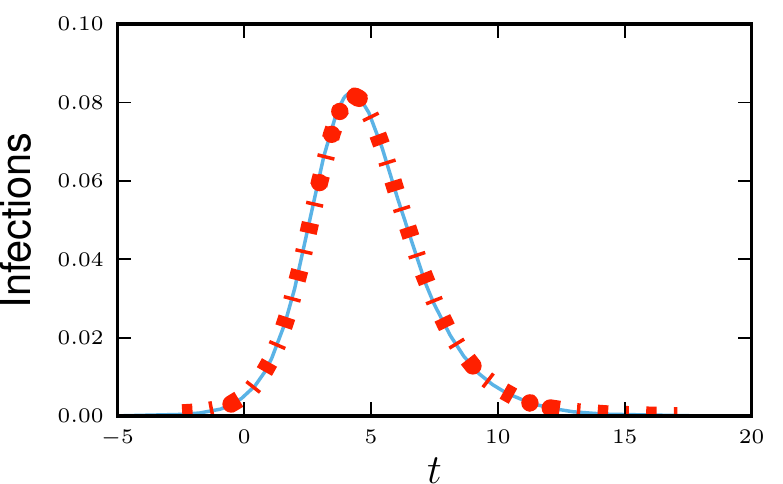}}
\caption{\footnotesize \textbf{Mixed Poisson Example (section~\ref{sec:MPex}).} Model predictions (dashed) match a simulated epidemic (solid) in a Mixed Poisson network with $5\times 10^5$ nodes.  The solid curve is a single simulation.  Time is chosen so that $t=0$ corresponds to $1$\% cumulative incidence.}
\label{fig:MPex}
\end{figure}
We consider a population whose distribution of expected degrees satisfies
\[
\rho(\kappa) = \begin{cases} \frac{1}{4} & 0\leq \kappa \leq 2\\
 \frac{1}{20} & 10\leq \kappa \leq 20\\
0 & \text{Otherwise}
\end{cases}
\]
Half the individuals have an expected degree between $0$ and $2$ uniformly, and the other half have expected degree between $10$ and $20$ uniformly.  This gives
\[
\Psi(x) =  \frac{1}{4(x-1)}\left( e^{2(x-1)} - 1 + \frac{e^{20(x-1)} - e^{10(x-1)}}{5} \right)
\]
We take $\beta = 0.15$ and $\gamma = 1$, and perform simulations with a population of size $5\times10^5$ generated using the algorithm of~\cite{miller:chunglu}.  We compare simulation and prediction in figure~\ref{fig:MPex}.

\subsection{Mean Field Social Heterogeneity}
\label{sec:ed_MFSH}
For the expected degree formulation of the Mean Field Social Heterogeneity (MFSH) model, the probability an edge exists between $u$ and $v$ at time $t$ is $\kappa_u\kappa_v/(N-1)\ave{K}$.  Whether this edge exists at one moment is independent of whether it exists at any other moment and what other edges exist.

\begin{figure}
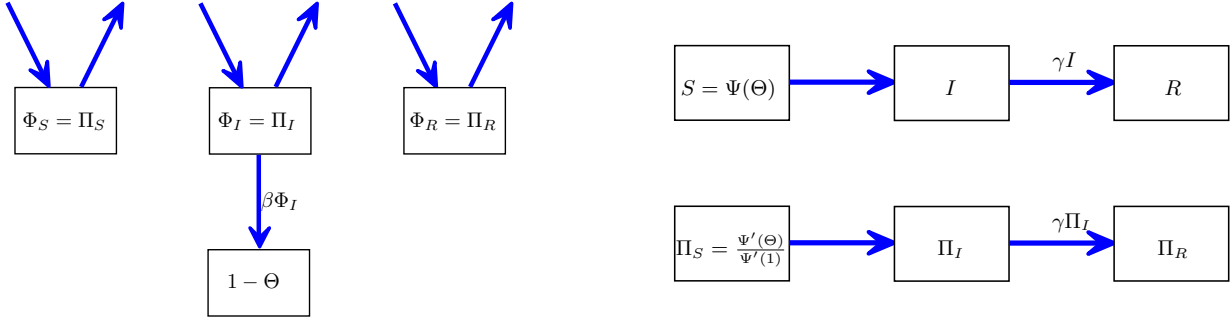

\parbox{\textwidth}{\parbox{0.4\textwidth}{\scalebox{0.8}{\begin{picture}(0,0)%
\includegraphics{edgeflux_AD_MFSH_all_flux.pdf}%
\end{picture}%
\setlength{\unitlength}{3947sp}%
\begingroup\makeatletter\ifx\SetFigFontNFSS\undefined%
\gdef\SetFigFontNFSS#1#2#3#4#5{%
  \reset@font\fontsize{#1}{#2pt}%
  \fontfamily{#3}\fontseries{#4}\fontshape{#5}%
  \selectfont}%
\fi\endgroup%
\begin{picture}(3883,2441)(2026,-3544)
\put(2609,-2047){\makebox(0,0)[b]{\smash{{\SetFigFontNFSS{10}{9.6}{\familydefault}{\mddefault}{\updefault}{\color[rgb]{0,0,0}$\Phi_S=\Pi_S$}%
}}}}
\put(4124,-2047){\makebox(0,0)[b]{\smash{{\SetFigFontNFSS{10}{9.6}{\familydefault}{\mddefault}{\updefault}{\color[rgb]{0,0,0}$\Phi_I=\Pi_I$}%
}}}}
\put(4094,-3310){\makebox(0,0)[b]{\smash{{\SetFigFontNFSS{10}{9.6}{\familydefault}{\mddefault}{\updefault}{\color[rgb]{0,0,0}$1-\Theta$}%
}}}}
\put(5661,-2047){\makebox(0,0)[b]{\smash{{\SetFigFontNFSS{10}{9.6}{\familydefault}{\mddefault}{\updefault}{\color[rgb]{0,0,0}$\Phi_R=\Pi_R$}%
}}}}
\put(4301,-2674){\makebox(0,0)[b]{\smash{{\SetFigFontNFSS{10}{9.6}{\familydefault}{\mddefault}{\updefault}{\color[rgb]{0,0,0}$\beta\Phi_I$}%
}}}}
\end{picture}%
}}\hfill \parbox{0.45\textwidth}{\scalebox{0.9}{\begin{picture}(0,0)%
\includegraphics{standardflux.pdf}%
\end{picture}%
\setlength{\unitlength}{3947sp}%
\begingroup\makeatletter\ifx\SetFigFontNFSS\undefined%
\gdef\SetFigFontNFSS#1#2#3#4#5{%
  \reset@font\fontsize{#1}{#2pt}%
  \fontfamily{#3}\fontseries{#4}\fontshape{#5}%
  \selectfont}%
\fi\endgroup%
\begin{picture}(3883,550)(1651,-1886)
\put(2038,-1700){\makebox(0,0)[b]{\smash{{\SetFigFontNFSS{9}{9.6}{\familydefault}{\mddefault}{\updefault}{\color[rgb]{0,0,0}$S=\Psi(\Theta)$}%
}}}}
\put(3589,-1700){\makebox(0,0)[b]{\smash{{\SetFigFontNFSS{9}{9.6}{\familydefault}{\mddefault}{\updefault}{\color[rgb]{0,0,0}$I$}%
}}}}
\put(5140,-1700){\makebox(0,0)[b]{\smash{{\SetFigFontNFSS{9}{9.6}{\familydefault}{\mddefault}{\updefault}{\color[rgb]{0,0,0}$R$}%
}}}}
\put(4382,-1509){\makebox(0,0)[b]{\smash{{\SetFigFontNFSS{9}{9.6}{\familydefault}{\mddefault}{\updefault}{\color[rgb]{0,0,0}$\gamma I$}%
}}}}
\end{picture}%
}\\[30pt]
\scalebox{0.9}{\begin{picture}(0,0)%
\includegraphics{standardflux.pdf}%
\end{picture}%
\setlength{\unitlength}{3947sp}%
\begingroup\makeatletter\ifx\SetFigFontNFSS\undefined%
\gdef\SetFigFontNFSS#1#2#3#4#5{%
  \reset@font\fontsize{#1}{#2pt}%
  \fontfamily{#3}\fontseries{#4}\fontshape{#5}%
  \selectfont}%
\fi\endgroup%
\begin{picture}(3883,550)(1651,-1886)
\put(2058,-1700){\makebox(0,0)[b]{\smash{{\SetFigFontNFSS{9}{9.6}{\familydefault}{\mddefault}{\updefault}{\color[rgb]{0,0,0}$\Pi_S=\frac{\Psi'(\Theta)}{\Psi'(1)}$}%
}}}}
\put(3589,-1700){\makebox(0,0)[b]{\smash{{\SetFigFontNFSS{9}{9.6}{\familydefault}{\mddefault}{\updefault}{\color[rgb]{0,0,0}$\Pi_I$}%
}}}}
\put(5140,-1700){\makebox(0,0)[b]{\smash{{\SetFigFontNFSS{9}{9.6}{\familydefault}{\mddefault}{\updefault}{\color[rgb]{0,0,0}$\Pi_R$}%
}}}}
\put(4432,-1509){\makebox(0,0)[b]{\smash{{\SetFigFontNFSS{9}{9.6}{\familydefault}{\mddefault}{\updefault}{\color[rgb]{0,0,0}$\gamma \Pi_I$}%
}}}}
\end{picture}%
}}}\\
\caption{\footnotesize \textbf{Mean Field Social Heterogeneity model}.  The flow diagram for a Mean Field Social Heterogeneity population (expected degree formulation).  This is similar to the actual degree formulation in figure~\ref{fig:mfsh}.  The new variables $\Pi_S$, $\Pi_I$, and $\Pi_R$ (bottom right) represent the probability that a newly formed edge connects to a susceptible, infected, or recovered node, they can be thought of as the relative rates that each group forms edges.  Since the test node does not cause infections, and the probability a contact is with a node of a given $\kappa$ is equal to the probability a new contact will be with a node of the given $\kappa$, the probability a current neighbor has a given state is equal to the probability a new neighbor has that state.  Thus each $\Phi$ variable equals the corresponding $\Pi$ variable.}
\label{fig:MFSH_ED}
\end{figure}

As before, we consider two nodes whose expected degrees differ by $\Delta \kappa$ and ask how much the additional $\Delta \kappa$ reduces the probability of being susceptible.  The definition of $\Theta$ is slightly more problematic here because having an edge at one moment is independent of having one later.  So it does not make sense to discuss the probability an edge did not transmit previously because the edge did not exist previously.  Instead, we note that in the MP case $(1-\Theta)\Delta \kappa$ could be interpreted as the probability that the additional amount of $\Delta \kappa$ ever contributed an edge that had transmitted infection.  Guided by this, we define $\Theta$ so that as $\Delta \kappa \to 0$, the extra amount of $\Delta \kappa$ has at some time contributed an edge that transmitted to the node with pobability $(1-\Theta)\Delta \kappa$.  This again leads to $\Psi(x)= \int_0^\infty e^{-\kappa(1-x)}\rho(\kappa)\, \mathrm{d}\kappa$.  The  flow diagram for individuals (figure~\ref{fig:MFSH_ED}) is unchanged:
\[
\dot{R} = \gamma I \, , \qquad\qquad  S = \Psi(\Theta)  \, , \qquad\qquad  I = 1 -S -R
\]

To find the evolution of $\Theta$, we define $\Phi_S$, $\Phi_I$, and $\Phi_R$ to be the probabilities a current edge connects $u$ to a susceptible, infected, or recovered node.  The probability that a small extra amount $\Delta \kappa$ currently contributes an edge and previously had a different edge that transmitted scales like $\Delta \kappa^2(1-\Theta)$.  Since $\Delta \kappa^2 \ll \Delta \kappa$, this is negligible compared to the probability that there is a current edge.  We conclude that at leading order, $\Phi_S\Delta\kappa$, $\Phi_I \Delta \kappa$, and $\Phi_R\Delta\kappa$ give the probability that the $\Delta \kappa$ contributes a current edge connected to a susceptible, infected, or recovered node and there has never been a transmission due to this extra $\Delta \kappa$.
 
We can construct a flow diagram between $\Phi_S\Delta \kappa$, $\Phi_I\Delta\kappa$, $\Phi_R\Delta\kappa$, and $(1-\Theta)\Delta \kappa$.  Because all of these have $\Delta \kappa$ in them, we factor it out to create to just use $\Phi_S$, $\Phi_I$, $\Phi_R$, and $1-\Theta$.  Because edges have no duration, there is no $\Phi_S$ to $\Phi_I$ or $\Phi_I$ to $\Phi_R$ flux (similar to the actual degree MFSH model).  Instead there is flux in and out of these compartments representing the continuing change of edges.  The $\Phi_I$ to $1-\Theta$ flux is $\beta \Phi_I$.

Because edges have no duration, the probability an edge connects to an individual of a given type is the probability a new edge connects to an individual of that type: $\Phi_S=\Pi_S$, $\Phi_I=\Pi_I$, and $\Phi_R=\Pi_I$ where $\Pi_S$, $\Pi_I$ and $\Pi_R$ are the probabilities that a newly formed edge connects to a susceptible, infected, or recovered node.\footnote{Unlike the actual degree formulation we do not need a factor of $\Theta$ in these because the smallness of $\Delta \kappa$ allows us to assume there has never been a previous transmission.}  We have $\Pi_S = \Psi'(\Theta)/\Psi'(1)$ and $\dot{\Pi}_R = \gamma \Pi_I$.  Since $\Pi_I = \Phi_I$, this means $\dot{\Pi}_R = -\gamma \dot{\Theta}/\beta$. Integrating gives $\Pi_R=\gamma(1-\Theta)/\beta$.  So $\Pi_I = 1-\Psi'(\Theta)/\Psi'(1)-\gamma(1-\Theta)/\beta$.  Since $\dot{\Theta} = -\beta \Phi_I=-\beta\Pi_I$, we finally have
\begin{align*}
\dot{\Theta} &= -\beta +\beta\frac{\Psi'(\Theta)}{\Psi'(1)} + \gamma(1-\Theta)\\
\dot{R} &= \gamma I \, , \qquad\qquad S = \Psi(\Theta) \, , \qquad\qquad I = 1-S-R
\end{align*}
Under appropriate limits the MFSH equations in $k$ reduce to those in $\kappa$ and vice versa, so the models are equivalent~\cite{miller:ebcm_hierarchy}.  Surprisingly, this system differs from the MP equations only in the first term of the $\dot{\Theta}$ equation.

\subsubsection{$\Ro$ and final size}
We find 
\[
\Ro=\frac{\beta}{\gamma} \frac{\ave{\hat{K}^2}}{\ave{K}}
\]
and the final size is $R(\infty) = 1- \Psi(\Theta(\infty))$ where $\Theta(\infty)$ solves
\[
\Theta = \frac{\beta}{\gamma}\left(1+\frac{\Psi'(\Theta)}{\Psi'(1)} \right) + 1
\]
Full details are in the Appendix.

\subsubsection{Example}
\label{sec:EDMFSHex}
\begin{figure}
\parbox{\textwidth}{\includegraphics{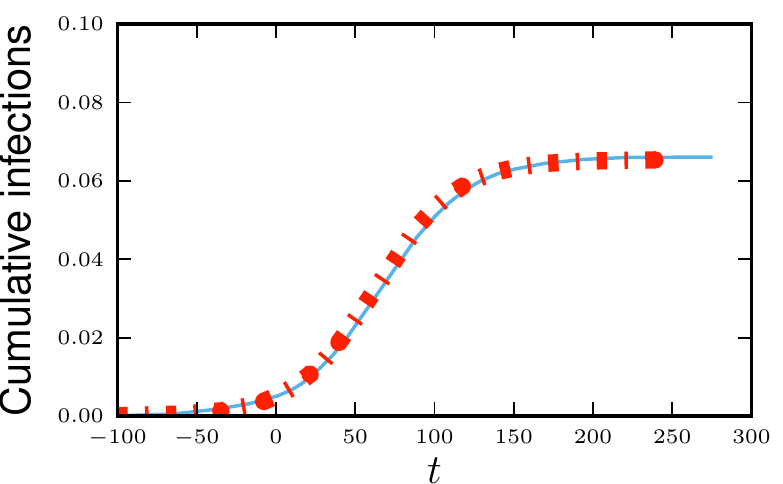}
\includegraphics{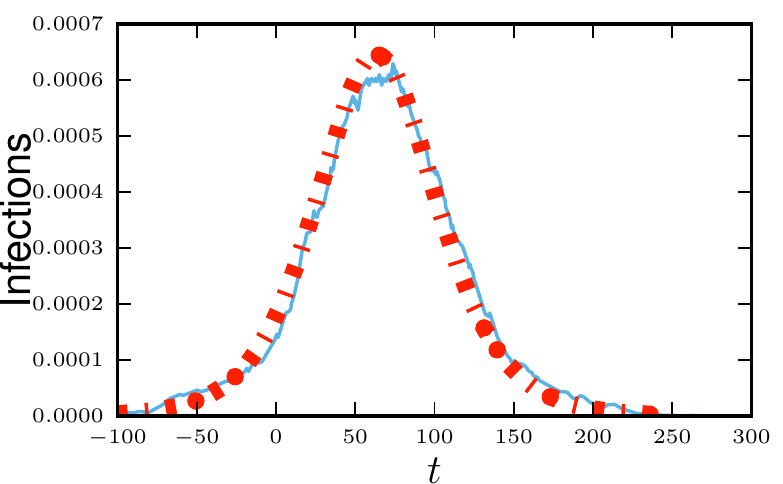}}
\caption{\footnotesize \textbf{Expected Degree MFSH Example (section~\ref{sec:EDMFSHex}).} Model predictions (dashed) match a simulated epidemic (solid) in a MFSH network with $15\times 10^6$ nodes.  The solid curve is a single simulation.  Time is chosen so that $t=0$ corresponds to $0.5$\% cumulative incidence.}
\label{fig:mfsh_ed_ex}
\end{figure}

For our example we take a population with $\rho(\kappa) = e^\kappa/(e^3-1)$ for $0<\kappa<3$ and $0$ otherwise, giving
\[
\Psi(x) = \frac{e^{3x}-1}{x(e^3-1)}
\]
We take $\gamma=1$ and $\beta=0.435$ and compare simulation with theory in figure~\ref{fig:mfsh_ed_ex}.  We choose these parameters to demonstrate that the approach remains accurate for small $\Ro=1.04$.  We use a population of $15\times 10^6$.  There is noise since the epidemic does not infect a large number of people.

\subsection{Dynamic Variable-Degree}
\label{sec:DVD}
The Dynamic Variable-Degree (DVD) model interpolates between the MP model and the expected degree formulation of the MFSH model.  Each node is assigned $\kappa$ using $\rho(\kappa)$ and creates edges at rate $\kappa \eta$ (joining to another node also creating an edge).  Existing edges break at rate $\eta$.  Thus a node has expected degree $\kappa$, though its value varies around $\kappa$.  In fact it is Poisson distributed over time.  

We define $\Theta$ such that the probability a small $\Delta \kappa$ has ever contributed an edge that has transmitted infection is $(1-\Theta)\Delta \kappa$.  We again have
\[
\dot{R} = \gamma I \, , \qquad\qquad  S = \Psi(\Theta)  \, , \qquad\qquad  I = 1 -S -R
\]
We generalize the earlier definitions and define $\Phi_S$, $\Phi_I$, and $\Phi_R$ to be the probabilities a current edge has never transmitted infection and connects to a susceptible, infected, or recovered node.

\begin{figure}
\parbox{\textwidth}{\parbox{0.5\textwidth}{\scalebox{0.9}{\begin{picture}(0,0)%
\includegraphics{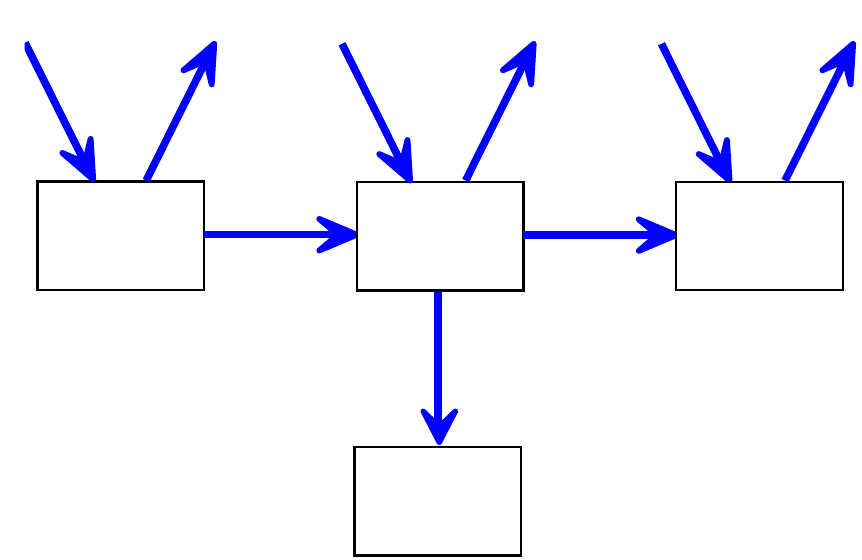}%
\end{picture}%
\setlength{\unitlength}{3947sp}%
\begingroup\makeatletter\ifx\SetFigFontNFSS\undefined%
\gdef\SetFigFontNFSS#1#2#3#4#5{%
  \reset@font\fontsize{#1}{#2pt}%
  \fontfamily{#3}\fontseries{#4}\fontshape{#5}%
  \selectfont}%
\fi\endgroup%
\begin{picture}(4120,2667)(1908,-2936)
\put(2495,-1457){\makebox(0,0)[b]{\smash{{\SetFigFontNFSS{9}{9.6}{\familydefault}{\mddefault}{\updefault}{\color[rgb]{0,0,0}$\Phi_S=\frac{\Psi'(\Theta)}{\Psi'(1)}$}%
}}}}
\put(4022,-1457){\makebox(0,0)[b]{\smash{{\SetFigFontNFSS{9}{9.6}{\familydefault}{\mddefault}{\updefault}{\color[rgb]{0,0,0}$\Phi_I$}%
}}}}
\put(5558,-1457){\makebox(0,0)[b]{\smash{{\SetFigFontNFSS{9}{9.6}{\familydefault}{\mddefault}{\updefault}{\color[rgb]{0,0,0}$\Phi_R$}%
}}}}
\put(4022,-2713){\makebox(0,0)[b]{\smash{{\SetFigFontNFSS{9}{9.6}{\familydefault}{\mddefault}{\updefault}{\color[rgb]{0,0,0}$1-\Theta$}%
}}}}
\put(2888,-359){\makebox(0,0)[b]{\smash{{\SetFigFontNFSS{9}{9.6}{\familydefault}{\mddefault}{\updefault}{\color[rgb]{0,0,0}$\eta\Phi_S$}%
}}}}
\put(3513,-359){\makebox(0,0)[b]{\smash{{\SetFigFontNFSS{9}{9.6}{\familydefault}{\mddefault}{\updefault}{\color[rgb]{0,0,0}$\eta\Pi_I$}%
}}}}
\put(4446,-359){\makebox(0,0)[b]{\smash{{\SetFigFontNFSS{9}{9.6}{\familydefault}{\mddefault}{\updefault}{\color[rgb]{0,0,0}$\eta\Phi_I$}%
}}}}
\put(5020,-359){\makebox(0,0)[b]{\smash{{\SetFigFontNFSS{9}{9.6}{\familydefault}{\mddefault}{\updefault}{\color[rgb]{0,0,0}$\eta\Pi_R$}%
}}}}
\put(5967,-359){\makebox(0,0)[b]{\smash{{\SetFigFontNFSS{9}{9.6}{\familydefault}{\mddefault}{\updefault}{\color[rgb]{0,0,0}$\eta\Phi_R$}%
}}}}
\put(1970,-359){\makebox(0,0)[b]{\smash{{\SetFigFontNFSS{9}{9.6}{\familydefault}{\mddefault}{\updefault}{\color[rgb]{0,0,0}$\eta\Pi_S$}%
}}}}
\put(4784,-1600){\makebox(0,0)[b]{\smash{{\SetFigFontNFSS{9}{9.6}{\familydefault}{\mddefault}{\updefault}{\color[rgb]{0,0,0}$\gamma\Phi_I$}%
}}}}
\put(4116,-2016){\makebox(0,0)[lb]{\smash{{\SetFigFontNFSS{9}{9.6}{\familydefault}{\mddefault}{\updefault}{\color[rgb]{0,0,0}$\beta\Phi_I$}%
}}}}
\end{picture}%
}}%
\hfill\parbox{0.475\textwidth}{\scalebox{0.95}{\begin{picture}(0,0)%
\includegraphics{standardflux.pdf}%
\end{picture}%
\setlength{\unitlength}{3947sp}%
\begingroup\makeatletter\ifx\SetFigFontNFSS\undefined%
\gdef\SetFigFontNFSS#1#2#3#4#5{%
  \reset@font\fontsize{#1}{#2pt}%
  \fontfamily{#3}\fontseries{#4}\fontshape{#5}%
  \selectfont}%
\fi\endgroup%
\begin{picture}(3883,550)(1651,-1886)
\put(2038,-1700){\makebox(0,0)[b]{\smash{{\SetFigFontNFSS{9}{9.6}{\familydefault}{\mddefault}{\updefault}{\color[rgb]{0,0,0}$S=\Psi(\Theta)$}%
}}}}
\put(3589,-1700){\makebox(0,0)[b]{\smash{{\SetFigFontNFSS{9}{9.6}{\familydefault}{\mddefault}{\updefault}{\color[rgb]{0,0,0}$I$}%
}}}}
\put(5140,-1700){\makebox(0,0)[b]{\smash{{\SetFigFontNFSS{9}{9.6}{\familydefault}{\mddefault}{\updefault}{\color[rgb]{0,0,0}$R$}%
}}}}
\put(4382,-1509){\makebox(0,0)[b]{\smash{{\SetFigFontNFSS{9}{9.6}{\familydefault}{\mddefault}{\updefault}{\color[rgb]{0,0,0}$\gamma I$}%
}}}}
\end{picture}%
}\\[30pt]
\scalebox{0.95}{\begin{picture}(0,0)%
\includegraphics{standardflux.pdf}%
\end{picture}%
\setlength{\unitlength}{3947sp}%
\begingroup\makeatletter\ifx\SetFigFontNFSS\undefined%
\gdef\SetFigFontNFSS#1#2#3#4#5{%
  \reset@font\fontsize{#1}{#2pt}%
  \fontfamily{#3}\fontseries{#4}\fontshape{#5}%
  \selectfont}%
\fi\endgroup%
\begin{picture}(3883,550)(1651,-1886)
\put(2058,-1700){\makebox(0,0)[b]{\smash{{\SetFigFontNFSS{9}{9.6}{\familydefault}{\mddefault}{\updefault}{\color[rgb]{0,0,0}$\Pi_S=\frac{\Psi'(\Theta)}{\Psi'(1)}$}%
}}}}
\put(3589,-1700){\makebox(0,0)[b]{\smash{{\SetFigFontNFSS{9}{9.6}{\familydefault}{\mddefault}{\updefault}{\color[rgb]{0,0,0}$\Pi_I$}%
}}}}
\put(5140,-1700){\makebox(0,0)[b]{\smash{{\SetFigFontNFSS{9}{9.6}{\familydefault}{\mddefault}{\updefault}{\color[rgb]{0,0,0}$\Pi_R$}%
}}}}
\put(4432,-1509){\makebox(0,0)[b]{\smash{{\SetFigFontNFSS{9}{9.6}{\familydefault}{\mddefault}{\updefault}{\color[rgb]{0,0,0}$\gamma \Pi_I$}%
}}}}
\end{picture}%
}}}  
\caption{\footnotesize \textbf{Dynamic Variable-Degree model.}  The flow diagrams for the DVD model.}
\label{fig:vd}
\end{figure}

We define $\Pi_S$, $\Pi_I$, and $\Pi_R$ to be the probabilities a new edge connects to a susceptible, infected, or recovered node.  We have $\Pi_S = \Psi'(\Theta)/\Psi'(1)$, \ $\Pi_I = 1-\Pi_S-\Pi_R$, and $\dot{\Pi}_R = \gamma \Pi_I$.  We build the flow diagram for $\Phi_S\Delta\kappa$, $\Phi_I\Delta\kappa$, $\Phi_R\Delta\kappa$, and $(1-\Theta)\Delta\kappa$.  There is flux into $\Phi_S\Delta\kappa$ at rate $\eta\Pi_S\Delta\kappa$ because this is the rate that the $\Delta \kappa$ leads to edge creation.  There is flux out of $\Phi_S\Delta \kappa$ at rate $\eta\Phi_S\Delta \kappa$ because existing edges break at rate $\eta$, and the probability such an edge exists is $\Phi_S\Delta \kappa$.   Similar fluxes exist for $\Phi_I$ and $\Phi_R$.  The flux out of $\Phi_I\Delta\kappa$ into $\Phi_R\Delta\kappa$ is $\gamma \Phi_I\Delta\kappa$ as before, and the flux into $(1-\Theta)\Delta \kappa$ is $\beta\Phi_I\Delta\kappa$.  We factor out $\Delta \kappa$ and the flow diagrams (figure~\ref{fig:vd}) are defined.

Because the existence of an edge from the test node $u$ to the neighbor $v$ has no impact on any other edges $v$ might have, the contacts $v$ has aside from $u$ are indistinguishable from the contacts of another node with the same $\kappa$, and so they are susceptible with the same probability:
$\Phi_S = \Pi_S$.  The fluxes into and out of $\Phi_S$ from edge creation/deletion balance, and the $\Phi_S$ to $\Phi_I$ flux is simply $-\dot{\Phi}_S$.  Using this and the other fluxes for $\Phi_I$, we conclude $\dot{\Phi}_I = -\dot{\Phi}_S + \eta \Pi_I - (\eta+\gamma+\beta)\Phi_I$.   Since $\dot{\Pi}_R=\gamma\Pi_I$ and $\dot{\Theta}=-\beta\Phi_I$, we can integrate this and find $\Phi_I =  -\Psi'(\Theta)/\Psi'(1) + \eta\Pi_R/\gamma + (\beta+\eta+\gamma)\Theta/\beta - (\eta+\gamma)/\beta$.  So $\dot{\Theta} = -\beta \Phi_I$ can be written in terms of $\Theta$ and $\Pi_R$.  We arrive at 
\begin{align}
\dot{\Theta} &= -\beta\Theta + \beta\frac{\Psi'(\Theta)}{\Psi'(1)} + \gamma(1- \Theta) + \eta\left(1- \Theta - \frac{\beta}{\gamma} \Pi_R\right) \, ,\\
\dot{\Pi}_R &= \gamma \Pi_I \, ,\qquad \Pi_S = \Psi'(\Theta)/\Psi'(1) \, ,\qquad \Pi_I = 1 - \Pi_S - \Pi_R \, , \\
\dot{R} &= \gamma I \, ,\quad\qquad S = \Psi(\Theta) \, ,\quad\qquad I=1-S-R \, .  
\end{align}
This is simpler than the DFD case because the smallness of $\Delta \kappa$ allowed us to assume that no previous transmission occurred.

\subsubsection{$\Ro$ and final size}
 We find 
 \[
 \Ro = \frac{\beta}{\beta+\eta+\gamma} \frac{\eta+\gamma}{\gamma} \frac{\ave{\hat{K}^2}}{\ave{K}} \, ,
 \]
The total proportion infected is $R(\infty) = 1-\Psi(\Theta(\infty))$ where
\[
\Theta(\infty) = \frac{\beta}{\beta+\eta+\gamma} \left(\frac{\eta+\gamma}{\gamma}\frac{\Psi'(\Theta(\infty))}{\Psi'(1)} + \frac{\eta+\gamma}{\beta} - \frac{\eta}{\gamma}\right) \, .
\]
Full details are in the Appendix.

\subsubsection{Example}
\label{sec:DVDex}
\begin{figure}
\parbox{\textwidth}{\includegraphics{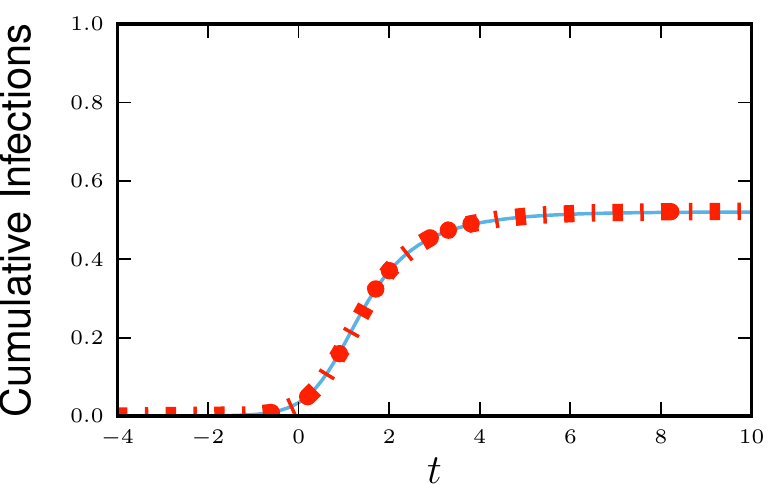} \includegraphics{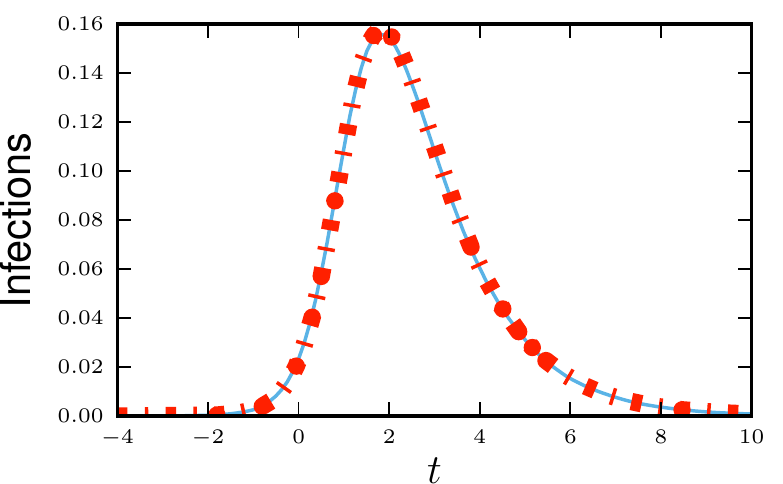}}
\caption{\footnotesize \textbf{Dynamic Variable-Degree Example (section~\ref{sec:DVDex}).} Model predictions (dashed) match the average of 92 simulated epidemics (solid) in a population of $10^4$ nodes.  For each simulation time is chosen so that $t=0$ corresponds to $3$\% cumulative incidence.  Then they are averaged to give the solid curve.}
\label{fig:varex}
\end{figure}

We choose the same distribution of $\kappa$ as of $k$ in the DFD example, $\NB(4,1/3)$.  We find
\[
\Psi(x) = \left(\frac{2}{3-e^{x-1}}\right)^4
\]
We take the same parameters $\beta=5/4$, \ $\gamma = 1$, and $\eta=1/2$.  Figure~\ref{fig:varex} shows that the equations accurately predict the spread.  As in the DFD and DC case, we use an average as the comparison point, taking 240 simulations and averaging the 92 resulting in epidemics.

The final size is larger than for the DFD model.  Although the average numbers of contacts are all the same, the increase in transmission routes when an individual had more contacts than expected outweighs the decrease when the number was less than expected.  The net effect is to increase the final size.

\section{Discussion}
We have introduced a new approach to study the spread of infectious diseases.  This edge-based compartmental modeling approach allows us to simultaneously consider the impacts of contact duration and social heterogeneity.  It is conceptually simple and leads to equations of comparable simplicity to the mass action model.  It produces a broad family of models which contains several known models as special cases.  It further allows us to investigate the effect of many behaviors which have previously been inaccessible to analytic study.

A significant contribution of this work is that it allows us to study the spread of a disease in a population for which some individuals have different propensity to form contacts while allowing us to explicitly incorporate the impact of contact duration.  The interaction of contact duration and numbers of overlapping contacts plays a significant role in the spread of many diseases, and in particular may play an important role in the spread of HIV.  These techniques open the door to studying these questions analytically rather than relying on simulation.

The edge-based compartmental modeling approach has a simple, graphical interpretation through flow diagrams.  This simplifies model description, and guides generalizations.  Because the derivations are straightforward, we need only track a small number of compartments, and we do not require any closure approximations, we propose that this is the ``correct'' perspective to study the deterministic dynamics of SIR epidemics on random networks.  Other existing techniques rely on approximations~\cite{eames:pair} or produce complicated or large systems of equations~\cite{volz:cts_time,lindquist,ball:network_eqns}.  This approach does not offer immediate insight into stochastic effects where methods such as branching processes are more appropriate.  In later papers, we use this approach to derive other generalizations, including different correlations in population structure, different disease dynamics, and populations that change structure in response to the spreading disease.

Our approach is limited by the assumption that infection of one neighbor of $u$ can be treated as independent of that of another neighbor.  This is a strong assumption, and prevents us from applying this model to SIS diseases for which individuals return to a susceptible state.  In this case, the assumption that $u$ does not infect its neighbors can alter the future state of $u$.  In the real population, if $u$ becomes infected, it can infect neighbors who then infect $u$ when $u$ returns to a susceptible state.  Thus if contact duration is nonzero, our predictions may be significantly altered.  This limitation is often not recognized but may lead to important failures of mean-field or mass action models when applied to a population for which contact duration is important~\cite{chatterjee:SIS}.

Treating neighbors as independent also means that if $v$ and $w$ are neighbors of $u$, we assume no alternate short path between $v$ and $w$ exists.  In particular, we assume that clustering~\cite{watts:collective,newman:clustering} is negligible: neighbors are unlikely to see one another.  This assumption applies equally to most existing analytic epidemic models, but it can be eliminated in special cases using techniques similar to those of~\cite{volz:clustered_result,miller:random_clustered,newman:cluster_alg,karrer:random_clustered}.

\section*{Acknowledgments}
JCM was supported by 1) the RAPIDD program of the Science and Technology Directorate, Department of Homeland Security and the Fogarty International Center, National Institutes of Health and 2) the Center for Communicable Disease Dynamics, Department of Epidemiology, Harvard School of Public Health under Award Number U54GM088558 from the National Institute Of General Medical Sciences.  Much of this work was a result of ideas catalyzed by the China-Canada Colloquium on Modeling Infectious Diseases in Xi'an, China, September 2009. EMV was supported by NIH K01 AI091440.  The content is solely the responsibility of the authors and does not necessarily represent the official views of the National Institute Of General Medical Sciences or the National Institutes of Health.

We thank S.~Bansal, M.~Lipsitch, R.~Meza, B.~Pourbohloul, P.~Trapman, and J.~Wallinga for useful conversations.

\appendix

\section{Appendix}
In this appendix, we give additional information for the edge-based compartmental modeling approach for the spread of susceptible-infected-recovered (SIR) diseases in different types of static and dynamic networks.  We give a more detailed discussion of the use of a test node and the assumption that test nodes do not cause infections.  We then discuss the calculation of $\Ro$, the behavior of our equations at early time (showing that the thresholds they predict are the same as those given by $\Ro$), and for some cases we give the final size prediction.
Finally, we show that the MFSH models we have used are in fact equivalent to some more familiar existing models.

\section{Selection of the test node}
The basis of our approach is the claim that the probability a randomly selected test node $u$ is susceptible, infected, or recovered is equal to the proportion of the population that is susceptible, infected, or recovered.  This claim implicitly assumes that the epidemic size grows deterministically: if stochastic effects could cause the outbreak to die out or even be slightly delayed, this claim is false.  The probability a random node $u$ is infected by time $t$ depends on whether an epidemic happens and, if so, how delayed it is.  So as in any ODE approach, our model is exact only once the outbreak is large enough to behave deterministically.

Our assumptions that the susceptible proportion of the population equals the probability $u$ is susceptible, the proportion infected equals the probability $u$ is infected, and the proportion recovered equals the probability $u$ is recovered allow us to move our focus away from the proportion in each state.  Instead we focus on the probability that $u$ is in each state.  Our goal remains to determine the course of the epidemic in the entire population, but our method will be to focus on the equivalent problem of finding the probability the randomly chosen test node $u$ has a given status.

In order to calculate the probability that $u$ is susceptible, infected, or recovered, we find another equivalent problem which is mathematically simpler.  We make a simplifying assumption which allows us to treat neighbors of $u$ as independent.  As it stands, if $w$ infects $u$, then $u$ can infect another neighbor $v$, meaning the satus of $v$ and $w$ are not independent.  We ignore this dependency, that is, we ignore transmissions from $u$ to its neighbors.  To make this mathematically precise, we prevent $u$ from transmitting infection to its neighbors.  This has no impact until after $u$ is infected, so it has no impact on the probability $u$ is susceptible.  It may affect the state of neighbors of $u$ once $u$ is infected, but it has no impact on the duration of infection of $u$, and so it does not alter the probability that $u$ is infected or recovered.  Consequently, this alteration of $u$ has no impact on the probability that $u$ is in any given state.  Thus our result for $S$, $I$, and $R$ is not affected by preventing $u$ from causing infection.

Consequently, we can calculate $S$, $I$, and $R$ as the probability that $u$ is susceptible, infected, or recovered under the assumption that $u$ is prevented from causing infection.  The result will give the proportion of the population that is susceptible, infected or recovered in the original epidemic.

\section{Simulation}
Both static networks and networks with mean field social heterogeneity satisfy the ``time homogeneity'' assumption of~\cite{kenah:second}.  That is, given the properties of $u$ and $v$, the \emph{a priori} probability that $u$ would transmit infection to $v$ if $u$ is infected is independent of the time at which $u$ becomes infected.  Consequently, for these cases we can use the Epidemic Percolation Network (EPN) approach of~\cite{kenah:EPN}.  In this, we consider each node $u$ in turn.  We assume that $u$ becomes infected and select the duration of infection from the appropriate exponential distribution of mean $\gamma$.  Given the duration of infection, for every node $v$ that $u$ might infect, we calculate the probability that $u$ infects $v$, and randomly determine whether $u$ infects $v$, and if so, how long it takes.  We then create a directed network by assigning edges from $u$ to each node it would infect with the edge weighted by the associated duration.  This directed network is an EPN.

To simulate an epidemic, we can choose a node to be the index case.  We then follow the epidemic as it passes from each node to the nodes that it would infect.  If the outbreak remains small, we discard it.  This can be done efficiently using Dijkstra's algorithm~\cite{dijkstra:algorithm}.  To quickly identify a node which sparks an epidemic, we can take the EPN and find the strongly-connected components within it.  Above the epidemic threshold there is a single giant strongly-connected component.  Any node from its ``in-component'' (including any node within the giant strongly-connected component) would spark an epidemic.  We choose any of these nodes randomly and use it as the index case.

The DVD, DFD, and DC models are harder to frame in terms of the EPN framework, so we use more traditional simulation techniques.   We use a Gillespie-style event-driven algorithm~\cite{gillespie:algorithm} and calculate whether the next event is a transmission, recovery, edge creation or edge breaking.  For the DFD model, edges break in pairs and neighbors are swapped.  These calculations are considerably slower because there are many events to track, only a few of which are directly relevant to disease transmission.

\section{$\Ro$, early growth, and final size}
In this section, we briefly turn away from the deterministic ODE methods and use branching process arguments to calculate $\Ro$ for each population.  We then return to the ODEs and linearize the equations about the equilibrium corresponding to a fully susceptible population.  We calculate the early growth rate, show that it is consistent with the branching process $\Ro$ above, and identify appropriate initial conditions.  Finally, for most of the models, we are able to calculate a final size relation.

The typically quoted definition of $\Ro$ is the number of new cases caused by a single randomly infected individual in a completely susceptible population.  However, a more careful definition is necessary in cases where the average individual in the population may have different properties than the average infected individual early in the epidemic.  The appropriate definition of $\Ro$ is the number of new cases an average infected individual causes early in an outbreak~\cite{diekmann:R0,trapman:analytical,miller:RSIcluster,volz:threshold}.  
In particular, for an epidemic on a network, a single node chosen randomly in the population and then infected will have (on average) $\ave{K}$ neighbors to infect, while early on the typical infected node has higher degree than a randomly chosen node and has at least one neighbor which is no longer susceptible.
Early in an outbreak, the probability mass function for a newly infected node in the actual degree case to have degree $k$ is $P_n(k)=kP(k)/\ave{K}$, while in the expected degree case the probability density function for a newly infected node to have expected degree $\kappa$ is $\rho_n(\kappa)=\kappa\rho(\kappa)/\ave{K}$.    Consequently, we must account for the fact that such a node has higher degree than average, but we must also account for the fact that such a node cannot infect the source of its infection. 

In our calculation of the early growth, we (as expected) find that if $\Ro<1$, the disease has negative growth rate.  We assume that the early growth is proportional to the leading eigenvector and use this to find appropriate initial conditions.  In practice this is unnecessary because effectively any appropriate initial condition (with almost all individuals and stubs being in a susceptible state) quickly converges to the leading eigenvector.  For our calculation of the final size, we are often able to identify a unique equilibrium corresponding to the state of the population after the disease has spread through.  For some models this is not possible.  As expected, if $\Ro<1$, we find that the only equilibrium corresponds to no large scale transmission, but if $\Ro>1$ there is an additional equilibrium which we can calculate to find the final size.

Most of our calculations for the early growth and final size are done under the assumption that an epidemic occurs and in the limit that the initial proportion infected goes to zero.  Thus our results are inappropriate for $\Ro<1$.  In calculating $\phi_S$ and $\phi_R$ in terms of $\theta$, we found that they take particular forms.  However, the imposed initial conditions could be different.  In the growing epidemic case, these early perturbations become insignificant as the number of infections becomes much larger than the initial conditions.  However, in the case of a decaying epidemic, the initial number of infections is always significant compared to the later number.  So the variations never disappear.  Thus if the initial conditions do not satisfy the formulae we derived, the later solution does not either.  This can still be handled using the edge-based compartmental modeling approach.  To correct for this in the CM model (and similar models) we would need to find the equation for $\dot{\phi}_S$ and $\dot{\phi}_I$ (resulting in a system more like the DFD equations).

\subsection{Actual degree models}

\subsubsection{CM}
\paragraph{$\Ro$}
In a CM network the expected number of infections a newly infected node causes is $\Ro= \sum_k P_n(k) (k-1)\beta/(\beta+\gamma)$ where $\beta/(\beta+\gamma)$ is the probability a node infects a neighbor prior to recovering.  The reason for the $k-1$ is that a newly infected node has one neighbor (its infector) who is not susceptible, and so there are $k-1$ susceptible neighbors.  So
\begin{align*}
\Ro &= \sum_k P_n(k) (k-1) \frac{\beta}{\beta+\gamma}\\
&= \sum_k \frac{k(k-1)P(k)}{\ave{K}} \frac{\beta}{\beta+\gamma}\\
&= \frac{\beta}{\beta+\gamma}\frac{\ave{K^2-K}}{\ave{K}}\\
&=\frac{\beta}{\beta+\gamma}\frac{\psi''(1)}{\psi'(1)}
\end{align*}
which is a well-known result for static CM networks.  This calculation is in agreement with previous results for CM networks~\cite{miller:heterogeneity,newman:spread,andersson:social,andersson:limit,trapman:analytical}.
 
In the special case of a network with a Poisson degree distribution, the probability of selecting a higher degree node and the reduction by one in the available number of susceptible neighbors exactly balance.  So for the Poisson distribution $\ave{K^2-K} = \ave{K}^2$ and $\Ro = \beta \ave{K}/(\beta+\gamma)$.  However,  this does not hold for more general distributions.  

\paragraph{Early Growth and Initial Conditions} 
We return to the deterministic equations
\begin{align*}
\dot{\theta} &=  -\beta\theta + \beta \frac{\psi'(\theta)}{\psi'(1)} + \gamma (1-\theta)  \, , \\
\dot{R} &= \gamma I  \, , \qquad\qquad  S = \psi(\theta)  \, , \qquad\qquad  I = 1 -S -R \, . 
\end{align*}
Clearly $\theta=1$ is an equilibrium solution corresponding to no transmission.  To test its stability, we linearize about $\theta=1$.  We set $\theta = 1+\epsilon$.  At leading order we find
\[
\dot{\epsilon} = \left(-\beta + \beta \frac{\psi''(1)}{\psi'(1)} - \gamma\right)\epsilon
\]
So at early times $\epsilon = Ce^{\lambda t}$ where
\[
\lambda = \beta \frac{\psi''(1)}{\psi'(1)} - (\beta+\gamma)
\]
The equilibrium loses stability as $\lambda$ transitions from negative to positive, $\beta\psi''(1)/\psi'(1)(\beta+\gamma)$, which is exactly the condition for $\Ro$ to transition from below $1$ to above $1$.  Both methods predict the same threshold.

To find appropriate initial conditions for $S$, $I$, and $R$, we could simply take $S=\psi(\theta)$, and choose any nonnegative $I$ and $R$ such that $1=S+I+R$.  As we solve forward, any error in $I$ and $R$ decays exponentially quickly.  If we wish to be more precise, we note that $\dot{I} = -\dot{S} - \gamma I$, and at leading order $\dot{S} = \dot{\theta}\psi'(\theta) = \lambda C e^{\lambda t} \psi'(1)$ to leading order.  We will have $I = Ke^{\lambda t}$, and we need to find $K$ in terms of $C$.  We get
$\lambda K e^{\lambda t}= -C \lambda \psi'(1)e^{\lambda t} -\gamma Ke^{\lambda t}$.  Solving this gives $K = -C\lambda\psi'(1)/(\gamma+\lambda)$, so the appropriate initial condition is
\[
\theta(0) = 1+C, \qquad S(0) = \psi(\theta(0)), \qquad I(0) = -\frac{C\lambda\psi'(1)}{\gamma+\lambda} , \qquad R(0) = 1-I(0)-S(0)
\]
where $C$ is a small, negative number.

However, in practice, there is no need to do this.  $I$ and $R$ have no role to play in determining $\theta$.  We simply require that $I+R = 1-\psi(\theta)$ initially.  Although our initial distribution of probability to $I$ and $R$ may differ from the true amount, it is a small effect initially and decays exponentially.  So in practice we can use any convenient assumption.

\paragraph{Final Size}
To calculate the final size, we note that as the epidemic dies out, the derivatives must all go to zero.  Thus we can set $\dot{\theta}=0$ and solve for $\theta(\infty)$.  Note that (if $\Ro>1$) this has two solutions, as there are two equilibrium conditions.  In one equilibrium the disease has not been introduced and $\theta=1$, while in the other the disease has spread and died out and $\theta<1$.  We want the smaller of the solutions, which corresponds to an epidemic occuring.
We solve
\[
\theta(\infty) = \frac{\gamma}{\beta+\gamma} + \frac{\beta}{\beta+\gamma} \frac{\psi'(\theta(\infty))}{\psi'(1)} 
\]
for the smaller solution.  In practice, this can be done by using a guess $\theta_1=0$, and then plugging $\theta_i$ into the right hand side to find $\theta_{i+1}$.  This iteration converges quickly, and if $\Ro>1$, the attracting solution is the solution we want.     The total fraction of the population infected in the course of an epidemic is $R(\infty)=1-\psi(\theta(\infty))$.

\subsubsection{Actual Degree MFSH}

\paragraph{$\Ro$}

To find $\Ro$ for the actual degree formulation of the MFSH model, we consider a newly infected node early in the epidemic.  The probability it has degree $k$ is $P_n(k)$.  Because it has a new set of neighbors at each moment, we do not have to account for the fact that it cannot infect the source of its infection, nor do we have to account for the fact that once it infects a neighbor, it cannot infect the neighbor again.  Thus at all times it has $k$ susceptible neighbors, so it causes new infections at rate $\beta k$ for the entire time it is infected.  The average duration of infection is $1/\gamma$, so the expected number of infections caused given $k$ is $\beta k/\gamma$.  Averaging over $k$, we have
\begin{align*}
\Ro &= \sum_k P_n(k) \frac{k\beta}{\gamma}\\
&= \frac{\beta}{\gamma} \sum \frac{k^2 P(k)}{\ave{K}}\\
&= \frac{\beta}{\gamma} \frac{\ave{K^2}}{\ave{K}}\\
&= \frac{\beta}{\gamma} \left(\frac{\psi''(1)}{\psi'(1)} + 1\right)
\end{align*}

\paragraph{Early Growth and Initial Conditions}
We begin with the equations
\begin{align*}
\dot{\theta} &= -\beta \theta + \beta \frac{\theta^2\psi'(\theta)}{\psi'(1)} - \theta\gamma \ln \theta\\
\dot{R} &= \gamma I  \, , \qquad\qquad  S = \psi(\theta)  \, , \qquad\qquad  I = 1-S-R
\end{align*}
we proceed similarly to the CM case.  We set $\theta = 1 +\epsilon$ and at leading order we find
\[
\dot{\epsilon} = \left(-\beta  + \beta \frac{2\psi'(1)+\psi''(1)}{\psi'(1)} - \gamma \right) \epsilon
\]
After some rearrangement, we have $\dot{\epsilon} = [\beta - \gamma + \beta\psi''(1)/\psi'(1)]\epsilon$.  So $\epsilon = Ce^{\lambda t}$ where 
\[
\lambda = \beta \frac{\psi''(1)}{\psi'(1)}  + \beta - \gamma
\]
The equilibrium loses stability exactly where $\Ro=1$.  The remaining calculations are identical to those of the CM case, and we find that the appropriate initial conditions are
as before except that the value of $\lambda$ is different
\[
\theta(0) = 1+C, \qquad S(0) = \psi(\theta(0)), \qquad I(0) = -\frac{C\lambda\psi'(1)}{\gamma+\lambda} , \qquad R(0) = 1-I(0)-S(0)
\]
(recall $C$ is a small, negative number).
As before, any reasonable initial condition with $\theta$ close to $1$, $S=\psi(\theta)$ and $I+R=1-S$ would be acceptable.

\paragraph{Final Size}
To find the final size of an epidemic, we set $\dot{\theta}$ to zero and solve for $\theta$.  We find
\[
\theta(\infty) = \exp\left[-\frac{\beta}{\gamma}\left(1-\frac{\theta(\infty)\psi'(\theta(\infty))}{\psi'(1)}\right)\right]
\]
If $\Ro>1$ this has two solutions, one with $\theta=1$, and one with $0<\theta<1$, which is the solution of interest.  Once this is found, the total fraction infected is $R(\infty) = 1 - \psi(\theta(\infty))$.

Note that if $\psi(x) = x^k$ for some $k$, this corresponds to the MFSH model with all individuals having the contact rate $k\beta$, which is the MA model and $\Ro = k \beta/\gamma$.  We find
\[
\theta = \exp \left( -\frac{\beta}{\gamma}[1-\theta^k]\right)
\]
Rewriting the left hand side as $\theta=S^{1/k}=(1-R(\infty))^{1/k}$ and raising both sides to the $k$'th power, we have
\[
1-R(\infty) = \exp \left(-\frac{k\beta}{\gamma} R(\infty) \right)
\]
Which is the well known final size relation for the MA model
\[
R(\infty) = 1-\exp(-\Ro R(\infty))
\]
\subsubsection{DFD}
\paragraph{$\Ro$}
To calculate $\Ro$ for this model, consider a randomly chosen newly infected node early in the epidemic.  It has degree $k$ with probability $P_n(k)$.  Initially this node has $k-1$ available susceptible neighbors.  Let us focus instead on the one edge from the infection source.  The stub may result in more infections if the edge is broken and reformed.  The probability that it breaks and reforms prior to recovering is $\eta/(\eta+\gamma)$.  The probability that it then causes infection prior to recovering is $\beta/(\beta+\gamma)$.  At this point the stub is connected to an infected neighbor, the same state it was at the beginning of infection and the process repeats.  So the probability this stub infects at least $n$ nodes is $r^n$ where $r=\eta\beta/[(\beta+\gamma)(\eta+\gamma)]$.  Summing this gives an expectation of $r/(1-r)$ new infections for this stub.  Now consider one of the $k-1$ stubs that are not the source of infection.  The probability that this stub transmits infection at least once is $\beta/(\beta+\gamma)$.  After this it is like the stub that received infection.  Thus the expected number of infections such a stub causes is $[\beta/(\beta+\gamma)][1+r/(1-r)]$ which can be rearranged into $[(\eta+\gamma)/\eta][r/(1-r)]$.

Adding these together, we see that a newly infected node is expected to cause
\begin{align*}
\Ro &= \sum_kP_n(k) \left[ \frac{r}{1-r} + (k-1)\frac{r(\eta+\gamma)}{\eta(1-r)}  \right]\\
&= \frac{r}{1-r} \sum_k \frac{kP(k)}{\ave{K}} \left[1 + (k-1) \frac{\eta+\gamma}{\eta}\right]\\
&= \frac{r}{1-r} \left( 1+\frac{\eta + \gamma}{\eta} \frac{\ave{K^2-K}}{\ave{K}} \right)\\
&= \frac{\beta\eta}{\gamma(\beta+\eta+\gamma)}\left( 1+\frac{\eta + \gamma}{\eta} \frac{\ave{K^2-K}}{\ave{K}} \right)\\
&= \frac{\beta}{(\beta+\eta+\gamma)}\left( \frac{\eta}{\gamma} + \frac{\eta + \gamma}{\gamma} \frac{\ave{K^2-K}}{\ave{K}}\right)\\
&= \frac{\beta}{(\beta+\eta+\gamma)}\left( \frac{\eta}{\gamma} + \frac{\eta + \gamma}{\gamma} \frac{\psi''(1)}{\psi'(1)}\right)
\end{align*}
where we have substituted for $r=\eta\beta/[(\beta+\gamma)(\eta+\gamma)]$.

\paragraph{Ealry Growth and Initial Conditions}

Our equations are
\begin{align*}
\dot{\theta} &= -\beta \phi_I \, ,\\
\dot{\phi}_S &= - \beta \phi_I \phi_S\frac{\psi''(\theta)}{\psi'(\theta)}  + \eta\theta\pi_S - \eta \phi_S\, ,  \\
\dot{\phi}_I &= \beta \phi_I \phi_S\frac{\psi''(\theta)}{\psi'(\theta)} + \eta \theta \pi_I - (\beta  + \gamma + \eta) \phi_I \, ,  \\
\dot{\pi}_R &= \gamma \pi_I \, , \qquad\qquad \pi_S = \frac{\theta\psi'(\theta)}{\psi'(1)} \, ,\qquad\qquad 
\pi_I = 1-\pi_R-\pi_S\, ,  \\ 
\dot{R} &= \gamma I \, ,\qquad\qquad S(t) = \psi(\theta) \, ,\qquad\qquad I(t) = 1-S-R \, . 
\end{align*}
Here we have a higher dimensional problem, and the equilibrium of interest is $\theta=1$, \ $\phi_S=1$, \ $\phi_I = 0$, \ $\pi_S=1$, and $\pi_I=0$.  We set $\theta = 1+\epsilon_1$, \ $\phi_S = 1 + \epsilon_2$, and $\phi_I = \epsilon_3$.  For $\pi_S$ we use $\pi_S = \theta\psi'(\theta)/\psi'(1)$.  For $\pi_I$, we set $\pi_I = \epsilon_4$ and use the fact that $\dot{\pi}_I = -\dot{\pi}_S - \gamma \pi_I$.  We linearize about the equlibrium.  We find
\begin{align*}
\dot{\epsilon}_1 &= - \beta \epsilon_3\\
\dot{\epsilon}_2 &= - \beta \epsilon_3 \frac{\psi''(1)}{\psi'(1)} + \eta \frac{2\psi'(1) + \psi''(1)}{\psi'(1)}\epsilon_1 - \eta \epsilon_2\\
\dot{\epsilon}_3 &= \beta \frac{\psi''(1)}{\psi'(1)}\epsilon_3  + \eta \epsilon_4- (\beta+\gamma+\eta)\epsilon_3\\
\dot{\epsilon}_4 &= \beta \frac{\psi'(1)+\psi''(1)}{\psi'(1)}\epsilon_3 - \gamma \epsilon_4
\end{align*}
which can be rewritten as the matrix equation 
\[
\diff{}{t} \begin{pmatrix}\epsilon_1\\\epsilon_2\\\epsilon_3\\\epsilon_4
  \end{pmatrix}
= \begin{pmatrix}
0 & 0&- \beta  &0 \\
\eta \frac{2\psi'(1) + \psi''(1)}{\psi'(1)} &- \eta & - \beta \frac{\psi''(1)}{\psi'(1)} &0\\
0 & 0 & \beta \frac{\psi''(1)}{\psi'(1)}  -(\beta+\gamma+\eta) & \eta \\
0 &0  & \beta \frac{\psi'(1)+\psi''(1)}{\psi'(1)}&- \gamma
\end{pmatrix}
\begin{pmatrix}\epsilon_1\\\epsilon_2\\\epsilon_3\\\epsilon_4
  \end{pmatrix}
\]
The standard solution technique for this is to find the largest eigenvalue of the matrix.    It is relatively straightforward to show that $0$ and $-\eta$ are always eigenvalues of this matrix.  The other two turn out to be the eigenvalues of the $2\times 2$ matrix forming the lower right corner.  They are
\[
\lambda_{1,2} = \frac{-(2\gamma+\beta+\eta-\beta\frac{\psi''(1)}{\psi'(1)}) \pm \sqrt{(2\gamma+\beta+\eta-\beta\frac{\psi''(1)}{\psi'(1)})^2 - 4\left(\gamma(\beta+\gamma+\eta)-\gamma\beta\frac{\psi''(1)}{\psi'(1)}-\eta\beta(1+\frac{\psi''(1)}{\psi'(1)}) \right)}}{2}
\]
For our above expression, if $\Ro=1$, then $\psi''(1)/\psi'(1) = (\gamma-\eta)/(\gamma+\eta) + \gamma/\beta$.  Placing this into our expression above, the largest eigenvalue becomes $0$.  If $\psi''(1)/\psi'(1)$ is larger than this threshold ($\Ro>1$) then an epidemic can occur.  If it is less than this threshold, this eigenvalue goes below zero, but there is still an eigenvector of zero whose eigenvector has $\epsilon_3$ and $\epsilon_4$ both zero (corresponding to $\phi_I$ and $\pi_I$ both zero).  If $\Ro$ is less than $1$, the values of $\epsilon_3$ and $\epsilon_4$ will decay according to the largest eigenvalue whose eigenvector has nonzero entries in the appropriate component.  

For initial conditions, if $\Ro>1$, we take $\lambda$ to be the largest eigenvalue, and the vector $(a,b,c,d)$ to be the corresponding eigenvalue.  Then the appropriate initial conditions are found by taking $\theta=1+\epsilon a$, \ $\phi_S = 1 + \epsilon b$, \ $\phi_I = \epsilon c$, and $\pi_I = \epsilon d$ where $\epsilon \ll 1$.  Note that we must choose our eigenvector such that $a<0$.  In practice however, so long as the initial amount of infection is taken to be very small, the initial conditions need not take this exact form; the solution will quickly converge to something of this form.  If $\Ro<1$, then the appropriate initial conditions come from a linear combination of the eigenvector of $0$ and the decaying eigenvector.  The coefficient of the $0$-eigenvector will be very small unless the initial introduction infected many individuals.


We note that in the previous cases, if $\Ro<1$, then the pertubation to $\theta$ decays, and $\theta$ returns to $1$.  Physically, this is an unrealistic result: it says that if transmission has already happened, then as time progresses, transmission is undone.  Here we do not see that.  If transmission has happened, then it does not decay, corresponding to the eigenvector of $0$.  The reason that this model is more correct, is that in the previous models, we found a relation between $\phi_I$ and $\theta$.  This relation implicitly assumes that the epidemic is growing.  If it is not growing, then this relation does not hold.  Eliminating the assumption from those models would result in additional equations for $\phi_S$ and $\phi_I$, and the system would look more like the DFD model.

\paragraph{Final Size}
We have not been able to find a simple expression for the final size of an epidemic in this case.  The system has multiple equilibria corresponding to possible states after the disease was introduced.  In the previous cases, we were able to find a closed form for the relation between $\theta$ and $\phi_I$.  The assumption made there was equivalent to stating that the early growth is dominated by the largest eigenvector.  In the previous cases, this assumption led to an analytic relation between $\phi_I$ and $\theta$.  In our case, we still want to make the equivalent assumption, which gives a constraint that determines which equilibrium is the final state.  However, we have not found a way to impose the constraint analytically.  Instead we must solve the system using initial conditions corresponding to a small number of cases to find the correct final size.  Thus we only have the final size as a numerical prediction.

\subsubsection{DC}

To calculate $\Ro$ for the DC model, we first define $r_a$, $r_d$, and $r_s$ to be the expected number of infections caused by a stub prior to recovery given that the stub is active and connected to a node other than the source, dormant, or active but connected to the source of infection at the time of infection.  

It is straightforward to show that if the stub is active and connected to a node other than the source, the probability that the edge transmits prior to breaking or recovery is $\beta/(\beta+\gamma+\eta_2)$.  The probability it breaks prior to recovery is $\eta_2/(\gamma+\eta_2)$.  Once it breaks, it is equivalent to a stub that was dormant at infection.  Thus $r_a = \beta/(\beta+\gamma+\eta_2) + \eta_2 r_d/(\gamma+\eta_2)$.

To find $r_d$, we note that a dormant stub must find a neighbor prior to recovery before it can cause any transmissions.  Once this happens, it is equivalent to a stub that was active at infection.  Thus $r_d = \eta_1r_a/(\gamma+\eta_1)$.  Combining this with our expression for $r_a$, we have
\[
r_a = \frac{\beta(\eta_1+\gamma)(\eta_2+\gamma)}{\gamma(\gamma+\eta_1+\eta_2)(\beta+\gamma+\eta_2)}.
\]

To find $r_s$, we note that infection cannot happen along that stub until the stub breaks and reforms at which point it is equivalent to an active stub, so $r_s = \eta_1\eta_2r_a/[(\gamma+\eta_1)(\gamma+\eta_2)]$.

The probability that a stub is active is $\xi=\eta_1/(\eta_1+\eta_2)$ and the probability it is dormant is $\pi=1-\xi$.  The total number of infections a node with degree $k_m$ is expected to cause is $r_s + (k-1)\xi r_a + (k-1)(1-\xi) r_d$.  Since the probability a newly infected node has degree $k_m$ is $P_n(k_m)=k_mP(k_m)/\ave{K_m}$, we find
\begin{align*}
\Ro &= \sum_{k_m} P_n(k_m) [(k_m-1)\xi r_a + (k_m-1)(1-\xi)r_d + r_s] \\
 &= \sum_{k_m} \frac{k_m P(k_m)}{\ave{K_m}} [(k_m-1)\xi r_a + (k_m-1)(1-\xi)r_d + r_s] \\
&= \sum_{k_m} \frac{k_m P(k_m)}{\ave{K_m}} \left[ (k_m-1)\left(\xi + (1-\xi) \frac{\eta_1}{\gamma+\eta_1}\right) + \frac{\eta_1\eta_2}{(\gamma+\eta_1)(\gamma+\eta_2)} \right]r_a\\
&= \left(\frac{\ave{K_m^2-K_m}}{\ave{K_m}} \frac{\eta_1}{\eta_1+\eta_2}\frac{\gamma+\eta_1+\eta_2}{\gamma+\eta_1} + \frac{\eta_1\eta_2}{(\gamma+\eta_1)(\gamma+\eta_2)}\right) r_a\\
&= \frac{\beta}{\gamma} \left[ \frac{\ave{K_m^2-K_m}}{\ave{K_m}} \frac{\eta_1}{\eta_1+\eta_2}\frac{\eta_2+\gamma}{\beta+\gamma+\eta_2}  + \frac{\eta_1\eta_2}{(\gamma+\eta_1+\eta_2)(\beta+\gamma+\eta_2)}\right]\\
&= \frac{\beta}{\beta+\eta_2+\gamma}\left(\frac{\ave{K_m^2-K_m}}{\ave{K_m}} \frac{\eta_1}{\eta_1+\eta_2}\frac{\eta_2+\gamma}{\gamma} + \frac{\eta_1\eta_2}{\gamma(\gamma+\eta_1+\eta_2)}\right)\\
&= \frac{\beta}{\beta+\eta_2+\gamma}\left(\frac{\psi''(1)}{\psi'(1)} \frac{\eta_1}{\eta_1+\eta_2}\frac{\eta_2+\gamma}{\gamma} + \frac{\eta_1\eta_2}{\gamma(\gamma+\eta_1+\eta_2)}\right)
\end{align*}

\paragraph{Early Growth and Initial Conditions}

We have not attempted to calculate the early growth rate because showing the details will not be particularly informative.  The method is similar to that for the DFD model.  If we wish to use appropriate initial conditions, we simply begin with $\theta$ approximately $1$, $\phi_S$ approximately $\pi$, $\phi_D$ approximately $\theta-\phi_S$, $\pi_S$ approximately $\pi$, and $\xi_S$ approximately $\xi$.  We can make all the $R$ variables $0$, and then set the $I$ variables to $I=1-S$, \ $\phi_I = \theta-\phi_S-\phi_D$, \ $\pi_I = \pi-\pi_S$, and $\xi_I = \xi-\xi_S$.  This will converge relatively quickly to the appropriate eigenvalue.  Alternately, we could solve the linear system and identify the appropriate eigenvalue and use it to find the initial conditions.

\paragraph{Final Size}
As in the DFD case, we need an additional constraint to identify the appropriate equilibrium.  We do not have this constraint analytically, so we must solve the ODE system numerically to find the final size.

\subsection{Expected Degree Models}
\subsubsection{MP}

\paragraph{$\Ro$}

We calculate $\Ro$ much as in the CM network.  We focus on all individuals with a given expected degree $\kappa$: these nodes have a Poisson degree distribution, and the fact that those with higher degree are more likely to become infected exactly cancels the reduction in available contacts, and so the expected number of remaining contacts of a newly infected node with expected degree $\kappa$ is $\kappa$.  So the expected number of infections such a node causes is $\kappa \beta/(\gamma+\beta)$.  To find $\Ro$, we must take a weighted average over the value of $\kappa$ for newly infected individuals.

The probability a newly infected individual has expected degree $\kappa$ is $\rho_n(\kappa)$.  So we find
\begin{align*}
\Ro &= \int_0^\infty \rho_n(\kappa) \frac{\kappa \beta}{\gamma+\beta}\\
&=\int_0^\infty \frac{\rho(\kappa)\kappa}{\ave{K}} \frac{\kappa \beta}{\gamma+\beta} \, \mathrm{d}\kappa\\
&= \frac{\ave{\hat{K}^2}}{\ave{K}} \frac{\beta}{\beta+\gamma}\\
&= \frac{\Psi''(1)}{\Psi'(1)} \frac{\beta}{\beta+\gamma}
\end{align*}
where $\ave{\hat{K}^2}$ denotes the average of $\kappa^2$.  It turns out $\ave{\hat{K}^2} = \ave{K^2-K}$, so this result is the same as the CM result.

\paragraph{Early Growth and Initial Conditions}

To calculate the early growth, we take
\begin{align*}
\dot{\Theta} &=  -\beta\Theta + \beta \frac{\Psi'(\Theta)}{\Psi'(1)} + \gamma (1-\Theta)  \, , \\
\dot{R} &= \gamma I  \, , \qquad\qquad  S = \Psi(\Theta)  \, , \qquad\qquad  I = 1 -S -R \, . 
\end{align*}
and set $\Theta = 1 +\epsilon$.  At leading order we have
\[
\dot{\epsilon} = \left(-\beta + \beta\frac{\Psi''(1)}{\Psi'(1)} - \gamma \right)\epsilon
\]
We find $\epsilon=Ce^{\lambda t}$ where
\[
\lambda = \beta\frac{\Psi''(1)}{\Psi'(1)}  - (\beta+\gamma)
\]
Looking at the threshold, we see that $\lambda = 0$ exactly where $\Ro=1$.  

To find appropriate initial conditions, we follow the CM case and find
\[
\Theta(0) = 1+C, \qquad S(0) = \Psi(\Theta(0)), \qquad I(0) = -\frac{C\lambda\Psi'(1)}{\gamma+\lambda} , \qquad R(0) = 1-I(0)-S(0)
\]
where $C$ is a small, negative number.

\paragraph{Final Size}

The final size of epidemics in MP networks can be calculated in much the same way as for CM networks.  We set $\dot{\Theta}=0$ and find
\[
\Theta(\infty) = \frac{\gamma}{\beta+\gamma} + \frac{\beta}{\beta+\gamma} \frac{\Psi'(\Theta(\infty))}{\Psi'(1)} 
\]
Then $S(\infty) = \Psi(\Theta(\infty))$ and $R(\infty)= 1-S(\infty)$.

\subsubsection{Expected Degree MFSH}

\paragraph{$\Ro$}

To find $\Ro$ for the actual degree formulation of the MFSH model, we consider a newly infected node early in the epidemic.  The probability density function for the expected degree $\kappa$ is $\rho_n(\kappa)$.  Because it has a new set of neighbors at each moment, we do not have to account for the fact that it cannot infect the source of its infection, nor do we have to account for the fact that once it infects a neighbor, it cannot infect the neighbor again.  Thus on average it has $\kappa$ susceptible neighbors, so it causes new infections at average rate $\beta \kappa$ for the entire time it is infected.  The average duration of infection is $1/\gamma$, so the expected number of infections caused given $k$ is $\beta \kappa/\gamma$.  Taking the average over all $\kappa$, we have
\begin{align*}
\Ro &= \int_0^\infty \rho_n(\kappa) \frac{\kappa\beta}{\gamma} \, \mathrm{d}\kappa\\
&= \frac{\beta}{\gamma} \int_0^\infty \frac{\kappa^2 \rho(\kappa)}{\ave{K}}\\
&= \frac{\beta}{\gamma} \frac{\ave{\hat{K}^2}}{\ave{K}}\\
&= \frac{\beta}{\gamma} \frac{\Psi''(1)}{\Psi'(1)}
\end{align*}

\paragraph{Early Growth and Initial Conditions}
Our governing equations are
\begin{align*}
\dot{\Theta} &= -\beta +\beta\frac{\Psi'(\Theta)}{\Psi'(1)} + \gamma(1-\Theta)\\
\dot{R} &= \gamma I \, , \qquad\qquad S = \Psi(\Theta) \, , \qquad\qquad I = 1-S-R
\end{align*}
Setting $\Theta = 1 + \epsilon$, we have
\[
\dot{\epsilon} = \left(\beta\frac{\Psi''(1)}{\Psi'(1)}-\gamma\right) \epsilon
\]
So $\epsilon = Ce^{\lambda t}$ where
\[
\lambda = \beta\frac{\Psi''(1)}{\Psi'(1)} - \gamma
\]
We see that the threshold for $\lambda=0$ is again the same as $\Ro=1$.

To find the initial conditions, we repeat our previous approach and find
\[
\Theta(0) = 1+C, \qquad S(0) = \Psi(\Theta(0)), \qquad I(0) = -\frac{C\lambda\Psi'(1)}{\gamma+\lambda} , \qquad R(0) = 1-I(0)-S(0)
\]
where $C$ is a small, negative number.

\paragraph{Final Size}
To find the final size we set $\dot{\Theta}=0$ and find that $\Theta(\infty)$ solves
\[
\Theta = \frac{\beta}{\gamma}\left(1+\frac{\Psi'(\Theta)}{\Psi'(1)} \right) + 1
\]
Then we have $R(\infty) = 1-\Psi(\Theta(\infty))$.

\subsubsection{DVD}

To calculate $\Ro$ for the DVD population, we begin by considering a newly infected node soon after disease is introduced.  Because nodes are infected with probability proportional to their expected degree, the probability density function for a node to have expected degree $\kappa$ given that it is newly infected is $\rho_n(\kappa)=\kappa\rho(\kappa)/\ave{K}$.  Given a newly infected node with expected degree $\kappa$, the expected number of additional neighbors (other than its infector) it has is also $\kappa$ (as in the static MP case).  For each of those neighbors, the probability that it transmits infection prior to recovering or breaking the edge is $\beta/(\beta+\eta+\gamma)$.  So the expected number of transmissions to neighbors it has when the infection occurs is $\kappa \rho(\kappa) \beta/[\ave{K}(\beta+\eta+\gamma)]$.

However, the node also has the opportunity to infect neighbors that it gains during its infectious period.  The probability that it creates a new edge before recovering is given by considering the recovery rate $\gamma$, and the edge creation rate $\kappa \eta$.  We track edge creations before recovery.  The probability that at least one edge creation occurs $\kappa \eta/(\gamma+\kappa \eta)$.  More generally, the probability that at least $n$ edge creations is $[\kappa \eta/(\gamma+\kappa \eta)]^n$.  If it gains at least $n$ neighbors, the probability that it infects the $n$-th neighbor before recovering or breaking the edge is $\beta/(\beta+\eta+\gamma)$.  So the probability that a node creates an $n$-th neighbor and infects that neighbor is $[\beta/(\beta+\eta+\gamma)][(\kappa\eta)/ (\gamma+\kappa+\eta)]^n$

The expected number of newly created neighbors which it infects can be found by summing the probability that a node creates and infects an $n$-th neighbor over all $n$.  This gives $[\beta/(\beta+\eta+\gamma)] \sum_n [\kappa \eta/(\gamma+\kappa\eta)]^n = [\beta/(\beta+\eta+\gamma)][\kappa\eta/\gamma]$.  
Adding the expected number of new and original neighbors infected together, the expected number of infections a node with $\kappa$ causes is $[\beta/(\beta+\eta+\gamma)]\kappa[1 + \eta/\gamma]$.  
Taking a weighted average over all $\kappa$ gives
\begin{align*}
\Ro &= \int_0^\infty \frac{\kappa\rho(\kappa)}{\ave{K}} \kappa\frac{\beta}{\beta+\eta+\gamma} \frac{\eta+\gamma}{\gamma} \, \mathrm{d}\kappa\\
&= \frac{\beta}{\beta+\eta+\gamma} 
\frac{\eta+\gamma}{\gamma} 
\frac{\ave{\hat{K}^2}}{\ave{K}} \\
&=\frac{\Psi''(1)}{\Psi'(1)}\frac{\beta}{\beta+\eta+\gamma}\frac{\eta+\gamma}{\gamma}
\end{align*}
The terms in this expression may be interpreted as follows: $\ave{\hat{K}^2}/{\ave{K}}$ gives the expected value of $\kappa$ for a newly infected node, $\beta/(\beta+\eta+\gamma)$ gives the probability that an edge which exists at any point during the infectious period will transmit infection prior to breaking or the infectious period ending, and $(\eta+\gamma)/\gamma = 1 + [\eta/\gamma]$ gives the expected number of susceptible contacts per expected degree to exist at infection ($1$) or be created prior to recovery ($\eta/\gamma$).  

\paragraph{Early growth}
We take the equations
\begin{align*}
\dot{\Theta} &= -\beta\Theta + \beta\frac{\Psi'(\Theta)}{\Psi'(1)} + \gamma(1- \Theta) + \eta\left(1- \Theta - \frac{\beta}{\gamma} \Pi_R\right) \, ,\\
\dot{\Pi}_R &= \gamma \Pi_I \, ,\qquad \Pi_S = \Psi'(\Theta)/\Psi'(1) \, ,\qquad \Pi_I = 1 - \Pi_S - \Pi_R \, , \\
\dot{R} &= \gamma I \, ,\quad\qquad S = \Psi(\Theta) \, ,\quad\qquad I=1-S-R \, .  
\end{align*}
We set $\Theta = 1+\epsilon_1$ and $\Pi_R = \epsilon_2$.  We note that $\dot{\Pi}_R = \gamma \Pi_I = \gamma(1-\Pi_S-\Pi_R)$.  At leading order we have
\begin{align*}
\dot{\epsilon}_1 &= -\beta \epsilon_1 + \beta \frac{\Psi''(1)}{\Psi'(1)} \epsilon_1 - \gamma\epsilon_1 + \eta\left(-\epsilon_1 - \frac{\beta}{\gamma}\epsilon_2\right)\\
\dot{\epsilon}_2 &= \gamma\left(-\frac{\Psi''(1)}{\Psi'(1)}\epsilon_1 - \epsilon_2\right)
\end{align*}
which becomes
\[
\diff{}{t}\begin{pmatrix} \epsilon_1\\\epsilon_2 \end{pmatrix}
= \begin{pmatrix}  \beta \frac{\Psi''(1)}{\Psi'(1)}-(\beta+\gamma+\eta) & -\frac{\eta\beta}{\gamma}\\
-\gamma \frac{\Psi''(1)}{\Psi'(1)} & -\gamma
\end{pmatrix}
\begin{pmatrix} \epsilon_1\\\epsilon_2 \end{pmatrix}
\]
The eigenvalues of a $2\times 2$ matrix solve $\lambda^2 - T\lambda + D$ where $T$ is the trace and $D$ the determinant.  So the dominant eigenvalue is
\[
\lambda = \frac{T + \sqrt{T^2-4D}}{2}
\]
If $T>0$, then the growth rate is positive.  To show that $T>0$ implies $\Ro>1$, note that $T>0$ implies $\beta\psi''(1)/\psi'(1) > \beta+\gamma+\eta$.  From this the product of the first two factors in our expression for $\Ro$ is greater than $1$.  Because $(\eta+\gamma)/\gamma>1$, it follows that $\Ro>1$.  If $T \leq 0$, then our equations predict growth if and only if $D<0$.  To complete our argument that the equations predict growth exactly when $\Ro>1$, we must show that if $T\leq 0$, then  $\Ro>1$ is equivalent to $D<0$.  We can show that
\[
D= -\beta (\eta+\gamma)\frac{\Psi''(1)}{\Psi'(1)} + \gamma (\beta+\gamma+\eta)
\]
From this a small amount of algebra shows $D<0$ is equivalent to $\Ro>1$.  Thus regardless of the sign of $T$, $\lambda>0$ exactly when $\Ro>1$, and conversely $\lambda<0$ exactly when $\Ro<1$.  So the predicted thresholds are the same.

To find the appropriate initial conditions, we can again take any sufficiently small reasonable initial condition and the particulars of the initial condition will be unimportant.  Alternately, we can note that the solution for $(\epsilon_1,\epsilon_2)$ must converge to $C e^{\lambda t} \mathbf{v}$ where $\mathbf{v}$ is the eigenvector of the eigenvalue $\lambda$.  This takes the value
\[
\mathbf{v} = \begin{pmatrix} \lambda + \gamma\\ \gamma \frac{\Psi''(1)}{\Psi'(1)}
\end{pmatrix}
\]
From this it is straightforward to find the appropriate initial conditions using the approaches seen before.

\paragraph{Final Size}
At the end of the epidemic, no infected nodes remain, and so $I(\infty)=\Phi_I(\infty)=\Pi_I(\infty)=0$.  We have $\Pi_R(\infty) = 1- \Pi_S(\infty) = 1- \Psi'(\Theta(\infty))/\Psi'(1)$.  Setting $\dot{\Theta}=0$ we find
\[
\Theta(\infty) = \frac{\beta}{\beta+\eta+\gamma} \left(\frac{\eta+\gamma}{\gamma}\frac{\Psi'(\Theta(\infty))}{\Psi'(1)} + \frac{\eta+\gamma}{\beta} - \frac{\eta}{\gamma}\right)
\]
We can solve this for $\Theta(\infty)$ using iterative methods.  The total fraction infected is 
\[
R=1-S=1-\Psi(\Theta(\infty))
\]

\section{Equivalence of MFSH models with pre-existing models}
The basic equations for the MFSH model used by other authors~\cite{andersonmay,may:hivdynamics, may:dynamics,moreno,pastor-satorras:scale-free} are
\begin{align*}
 \dot{S}_k &= -\beta kS_k \zeta\\
 \dot{I}_k &= \beta kS_k \zeta - \gamma I_k\\
 \zeta &= \frac{\sum_k kP(k) I_k}{\ave{K}}
 \end{align*}
However, in the actual degree case we have derived
\begin{align}
\dot{\theta} &= -\beta \theta + \beta \frac{\theta^2\psi'(\theta)}{\psi'(1)} - \theta\gamma \ln \theta\\
\dot{R} &= \gamma I  \, , \qquad\qquad  S = \psi(\theta)  \, , \qquad\qquad  I = 1-S-R
\end{align}
It is not immediately obvious that these are equivalent.  To see that they are, we first reduce the dimensions of the first system.  We note that the equation for $\dot{S}_k$ has as solution 
\[
S_k = e^{-\beta k \int_{-\infty}^t \zeta(t') \, \mathrm{d}t'}
\]
We set $\alpha = e^{-\beta \int_{-\infty}^t \zeta(t') \, \mathrm{d}t'}$ and then $S_k = \alpha^k$.  Our goal is to show that in fact, $\alpha$ solves the same equation as $\theta$.
We begin by noting that 
\[
\dot{\alpha} = -\beta \zeta \alpha
\]
So  $\zeta = -\dot{\alpha}/\beta\alpha$

We now move to finding $\dot{\zeta}$.
\begin{align*}
\dot{\zeta} &= \frac{\sum_k kP(k) \dot{I}_k}{\ave{K}}\\
&= \frac{ \sum_k k P(k)  [ \beta k S_k \zeta - \gamma I_k]}{\psi'(1)}\\
&= \beta\zeta\frac{\sum_k k^2 P(k) \alpha^k \zeta}{\psi'(1)} - \frac{\sum_k k P(k) \gamma I_k}{\psi'(1)}\\
&= \beta\zeta\frac{\sum_k (k^2-k+k) P(k) \alpha^k \zeta}{\psi'(1)} - \gamma \zeta\\
&= \beta\zeta\frac{ \psi''(\alpha)\alpha^2 + \psi'(\alpha)\alpha}{\psi'(1)} - \gamma \zeta\\
&= \alpha\beta\zeta\frac{  \psi''(\alpha) \alpha+ \psi'(\alpha)}{\psi'(1)} - \gamma \zeta\\
&= \alpha\beta\zeta\frac{  \diff{}{\alpha}(\alpha\psi'(\alpha))}{\psi'(1)} - \gamma \zeta
\end{align*}
We substitute $\zeta = - \dot{\alpha}/\beta\alpha$ to express this as a derivative.
\begin{align*}
\dot{\zeta} &= -\dot{\alpha}\frac{\diff{}{\alpha}(\alpha \psi'(\alpha))}{\psi'(1)} + \frac{\gamma}{\beta} \frac{\dot{\alpha}}{\alpha}\\
&= \diff{}{t} \left [ -\frac{\alpha \psi'(\alpha)}{\psi'(1)} + \frac{\gamma}{\beta} \ln \alpha\right]
\end{align*}
We can integrate this to find
\[
\zeta =  1 - \frac{\alpha \psi'(\alpha)}{\psi'(1)} + \frac{\gamma}{\beta} \ln \alpha
\]
(using the fact that $\zeta\to 0$ and $\alpha\to 1$ at early time) and so $\dot{\alpha} = -\beta \alpha\zeta$
becomes
\[
\dot{\alpha} = - \beta \alpha + \beta \alpha^2 \frac{\psi'(\alpha)}{\psi'(1)} - \alpha \gamma \ln \alpha
\]
which means that $\alpha$ solves the same equation as $\theta$ for the fixed degree version of the MFSH equations.  Since $S_k = \alpha^k$ is the same formula as we would find for $S_k$ in terms of $\theta$, this shows that in fact the two systems of equations are equivalent.

We are not the first to see that the usual system can be simplified into a handful of equations, but the approach we have used to derive these equations is new.  Previous authors have simply observed that the $S_k$ equation can be solved, done so, and then used a change of variables.  The resulting equations are equivalent to our own, but are written in terms of slightly different variables.  The advantage of our system is that the variables connect more easily to meaningful quantities, so it can be derived directly, and it can be related to the other edge-based compartmental models.

The usual model can be altered to allow for continuous contact rates, which would yield
\begin{align*}
 \dot{S}_\kappa &= -\beta \kappa S_\kappa \zeta\\
 \dot{I}_\kappa &= \beta \kappa S_\kappa \zeta - \gamma I_\kappa\\
 \zeta &= \frac{\int_0^\infty \kappa \rho(\kappa) I_\kappa \, \mathrm{d}\kappa}{\ave{K}} 
 \end{align*}
A similar approach shows that this is equivalent to our expected degree formulation of the MFSH equations.

\bibliographystyle{plain}
\small
\bibliography{paper1}

\end{document}